\documentclass[12pt]{iopart}

\usepackage{iopams}

\usepackage{amssymb,mathrsfs}
\usepackage{graphicx}

\newcommand{\prt}{\partial}
\newcommand{\sn}{\mathrm{sn}}
\newcommand{\cn}{\mathrm{cn}}

\newcommand{\la}{\lambda}

\begin{document}

\title[DNLS equation theory]
{Evolution of initial discontinuities in the DNLS equation theory}

\author{A M Kamchatnov}

\address{Institute of Spectroscopy, Russian Academy of Sciences,
Moscow, Troitsk, 108840, Russia}
\ead{kamch@isan.troitsk.ru}

\begin{abstract}
We present the full classification of wave patterns evolving from an initial step-like
discontinuity for arbitrary choice of boundary conditions at the discontinuity location
in the DNLS equation theory. In this {\it non-convex dispersive hydrodynamics} problem,
solutions of the Whitham modulation equations are mapped to parameters of a modulated
wave by two-valued functions what makes situation much richer than that for a convex
case of the NLS equation type. In particular, new types of simple-wave-like structures
appear as building elements of the whole wave pattern. The developed here theory can find
applications to propagation of light pulses in fibers and to the theory of Alfv\'en
dispersive shock waves.
\end{abstract}

\submitto{\JPA}


%\maketitle

\section{Introduction}\label{sec1}

The problem of classification of wave structures evolving from initial discontinuities has played
important role since the classical paper
of B.~Riemann \cite{riemann}. Complemented by the jump conditions of W.~Rankine \cite{rankine} and
H.~Hugoniot \cite{hugoniot-1,hugoniot-2}, it provided a prototypical example of formation of shocks
in dispersionless media with small viscosity, and the full classification of possible wave patterns
evolving from initial discontinuities with general initial data in adiabatic flows of ideal gas was obtained by
N.~Kotchine \cite{kochin}. These results were generalized to the class of so-called {\it genuinely
nonlinear hyperbolic} systems (see, e.g., \cite{lax-2006,dafermos}), however, situation beyond this class is
much more complicated and suffers from ambiguity of possible solutions. One of the methods to remove
this ambiguity is introduction of small viscosity into equations followed by taking the limit of zero
viscosity. This approach seems very natural from physical point of view since it provides some information
on the inner structure of viscous shocks. At the same time, there exists another method of regularization
of hydrodynamics-like equations, namely, introduction of small dispersion. Although in this case the limit
of zero dispersion does not lead to the same shock structure, this approach is of considerable interest
since, on one side, it is related with the theory of dispersive shock waves (DSWs) that finds a number of
physical applications (see, e.g., review article \cite{eh-16} and references therein) and, on another hand, there are situations
when the regularized equation belongs to the class of {\it completely integrable} equations and therefore
it admits quite thorough investigation including even cases of {\it non-genuinely nonlinear hyperbolic} systems.

The simplest example of dispersive nonlinear evolution equation is apparently the famous KdV equation
and in this case the solution of the Riemann problem is extremely simple: A.~V.~Gurevich and L.~P.~Pitaevskii
showed \cite{gp-73} with the use of Whitham modulation theory \cite{whitham-65} that there are only two possible
ways of evolution of initial discontinuity---it can evolve into either rarefaction wave or DSW whose
parameters can be expressed in explicit analytical form by solving the Whitham equations. This
result was obtained without explicit use of the complete integrability of the KdV equation \cite{ggkm},
but its extension to the NLS equation became possible \cite{eggk-95} only after derivation of the Whitham
modulation equations \cite{FL-86,pavlov-87} by the methods based on the inverse scattering transform
for the NLS equation \cite{zs-71} which means its complete integrability. It was shown in Ref.~\cite{eggk-95}
that the NLS equation evolution of any initial discontinuity leads to a wave pattern consisting of a sequence
of building blocks two of which are represented by either the rarefaction wave or the DSW, and they are separated
by plateau, or vacuum, or two-phase self-similar solution close to unmodulated nonlinear periodic wave. The
rarefaction waves are here self-similar simple wave solutions of the dispersionless limit of the NLS
equation (i.e., of the shallow water equations) and DSW is described by a self-similar solution of the Whitham
modulation equations. In total, there are six different possible wave patterns that can evolve from a
given initial discontinuity. Similar classification of wave patterns was also established for the
dispersive shallow water Kaup-Boussinesq equation \cite{egp-01,cikp-17}.

For classification of wave patterns arising in solutions of the Riemann problem of the KdV or NLS type, it
is important that the corresponding dispersionless limits (Hopf equation or shallow water equations) are
represented by the genuinely nonlinear hyperbolic equations. If it is not the case, then the classification of the
KdV-NLS type becomes insufficient and it was found that it should include new elements---kinks or trigonometric
dispersive shocks---for mKdV \cite{marchant} and  Miyata-Camassa-Choi \cite{ep-11} equations. The mKdV
equation is a modification of KdV equation and it also describes a unidirectional propagation of wave
with a single field variable, so it can be considered as a simplest example of {\it non-convex dispersive
hydrodynamics}. In spite of its relative simplicity, the full classification of the wave patterns in the
solution of the Riemann problem is much more complicated than that in the KdV equation case and it was achieved
in Ref.~\cite{kamch-12} for the Gardner equation (related with the mKdV equation) with the use of Riemann
invariant form of the Whitham modulation equations obtained in Ref.~\cite{pavlov-95}. These results were adapted
to mKdV equation in Ref.~\cite{ehs-17} and for this equation the Whitham modulation equations were obtained
by the direct Whitham method in Ref.~\cite{don-75}. Instead of two possible patterns in KdV case, in the
mKdV-Gardner case we have eight possible wave structures which depend now not only on the sign of the
jump at the discontinuity, but also on the values of wave amplitudes at both its sides. No similar classification
has been obtained yet for two-directional waves although important partial results were obtained in Ref.~\cite{ep-11}
for the Miyata-Camassa-Choi equation.
However, this equation is not completely integrable and although the principles of such a
classification are the same for completely integrable and non-integrable equations, we prefer here to turn
first to the case of completely integrable {\it derivative nonlinear Schr\"odinger} (DNLS) equation when more
complete study is possible.

Thus, in this paper, we shall give full solution of the Riemann problem for evolution of initial discontinuities in
the theory of the DNLS equation
\begin{equation}\label{DNLS}
  \rmi\Psi_t+\frac12\Psi_{xx}-\rmi(|\Psi|^2\Psi)_x=0.
\end{equation}
This equation appears in the theory of nonlinear Alfv\'en waves in plasma physics (see, e.g., \cite{kbhp-88}
and references therein) and in nonlinear optics (see, e.g., \cite{avc-86} and references therein). Its
complete integrability was established in \cite{kn-78,wki-79}, periodic solution and Whitham modulation
equations were derived in \cite{kamch-90a,kamch-90b}. Partial solution of the Riemann problem was obtained
in Ref.~\cite{gke-92}, however, only in the sector of the NLS equation type structures. Here we develop the method
which permits one to predict a wave pattern arising from any given data for an initial discontinuity.
The method is quite general and it was applied to the generalized NLS equations \cite{ik-17} with Kerr-type
cubic nonlinearity added to (\ref{DNLS}), what is important for nonlinear optics applications, and to
the Landau-Lifshitz equation for magnetics with easy-plane anisotropy \cite{ikcphi-17}. Here we develop
a similar theory for the equation (\ref{DNLS}).

\section{Hydrodynamic form of the DNLS equation and dispersion law for linear waves}\label{sec2}

In many situations, it is convenient to transform the DNLS equation (\ref{DNLS}) to the so-called
{\it hydrodynamic form} what is achieved by means of the substitution
\begin{equation}\label{304.2}
  \Psi(x,t)=\sqrt{\rho(x,t)}\exp\left(\rmi\int^xu(x',t)\rmd x'\right).
\end{equation}
After separation of real and imaginary parts, this equation is easily reduced to the system
\begin{eqnarray}\label{304.2a}
   \rho_t+\left[\rho\left(u-\frac32\rho\right)\right]_x=0,\\
   \label{304.2b}
   u_t+uu_x-(\rho u)_x+\left(\frac{\rho_x^2}{8\rho^2}-\frac{\rho_{xx}}{4\rho}\right)_x=0.
\end{eqnarray}
These equations can be interpreted as hydrodynamic form of the DNLS equation with Eq.~(\ref{304.2a})
playing the role of the continuity equation and Eq.~(\ref{304.2b}) of the Euler equation for a fluid
with depending on the flow velocity ``pressure'' $\rho u$ and ``quantum pressure'' represented by the last term.
However, one should keep in mind that we are dealing with an anisotropic medium where the flux of mass in
(\ref{304.2a}) does not coincide with the momentum density. As a result, the conservation of momentum
equation takes the form
\begin{equation}\label{305.10}
  [\rho(u-\rho)]_t+\left[\rho u^2-3u\rho^2+2\rho^3-\frac14\rho(\ln\rho)_{xx}\right]_x=0.
\end{equation}
This feature of the DNLS equation, which in our case means that the `right' and `left' directions of
wave propagation cannot be exchanged by an inversion operation $x\to -x$, can be illustrated by the
linear approximation.

Let us consider linear waves propagating along the background flow $(\rho_0,u_0)$, that is
$\rho=\rho_0+\rho'$, $u=u_0+u'$, where $|\rho'|\ll\rho_0$, $|u'|\ll u_0$. Linearization with respect to
small variables $\rho',\,u'$ yields the system
\begin{equation}\label{304.3}
\eqalign{
\rho'_t+(u_0-3\rho_0)\rho'_x+\rho_0u'_x=0 \cr
u'_t+(u_0-\rho_0)u'_x-\frac1{4\rho_0}\rho'_{xxx}=0.
}
\end{equation}
Looking for the plane wave solution $\rho',u'\propto\exp[\rmi(kx-\omega t)]$, we find that it exists if only
the dispersion law
\begin{equation}\label{304.4}
  \omega(k)=k\left[u_0-2\rho_0\pm\sqrt{\rho_0(\rho_0-u_0)+k^2/4}\right]
\end{equation}
is fulfilled. In the limit of small wave vectors $k$ we find
\begin{equation}\label{305.11}
  \omega(k)\approx \left(u_0-2\rho_0\pm\sqrt{\rho_0(\rho_0-u_0)}\right)k\pm\frac{k^3}{8\sqrt{\rho_0(\rho_0-u_0)}}.
\end{equation}
As we see, there are two modes of propagation of linear waves with different absolute values of propagation
velocities even for medium at rest with $u_0=0$: the initial disturbance decays to two wave packets propagating
with different absolute values of group velocities.

Another important feature of the dispersion law (\ref{304.4}) is that it leads to modulationally unstable
modes with complex $\omega$ for $u_0>\rho_0$. In this paper, we shall confine ourselves to the stable
situations only.

The above properties of the wave propagation in the DNLS equation theory are preserved in the weakly
nonlinear cases, that is if we take into account
weak nonlinear effects in the above modes with $\rho'$ small but finite.
Before proceeding to this task, we shall consider in the next section the dispersionless
dynamics when the dispersion effects are completely neglected.

\section{Dispersionless limit}\label{sec3}

The nonlinear and dispersive effects have the same order of magnitude, when in Eqs.~(\ref{304.2a}), (\ref{304.2b})
we have $u^2\sim \rho u\sim\rho_{xx}/\rho$, hence the last term in Eq.~(\ref{304.2b}) can be neglected
if the variables $\rho$ and $u$ change little on distances $\Delta x\sim1/\rho$.
In this dispersionless approximation, the flow is governed by the equations
\begin{equation}\label{305.7}
  \eqalign{
  \rho_t+\left[\rho\left(u-\frac32\rho\right)\right]_x=0,\cr
  u_t+uu_x-(\rho u)_x=0
  }
\end{equation}
or
\begin{equation}\label{306.13c}
   \left(
   \begin{array}{c}
   \rho \\
    u \\
   \end{array}
   \right)_t
       +\mathbb{A}\left(
        \begin{array}{c}
         \rho \\
           u \\
         \end{array}
            \right)_x=0,\quad
             \mathbb{A}=\left(
              \begin{array}{cc}
                 u-3\rho & \rho \\
                     -u & u-\rho \\
              \end{array}
              \right).
\end{equation}
The characteristic velocities of this system
\begin{equation}\label{305.4}
  v_{\pm}=u-2\rho\pm\sqrt{\rho(\rho-u)}
\end{equation}
coincide, naturally, with the phase velocities $\left.\omega/k\right|_{k\to0}$ for the dispersion
laws (\ref{304.4}) in the long wave limit.
The system (\ref{306.13c}) of first-order equations can be easily transformed to a diagonal form
\begin{equation}\label{306.17}
\frac{\prt r_+}{\prt t}+v_+\frac{\prt r_+}{\prt x}=0,\quad
\frac{\prt r_-}{\prt t}+v_-\frac{\prt r_-}{\prt x}=0
\end{equation}
for the Riemann invariants
\begin{equation}\label{306.15}
  r_+=u/2-\rho+\sqrt{\rho(\rho-u)},\quad  r_-=u/2-\rho-\sqrt{\rho(\rho-u)},
\end{equation}
with the velocities (\ref{305.4}) expressed in terms of the Riemann invariants as
\begin{equation}\label{306.16}
  v_+=\frac32r_++\frac12r_-,\quad v_-=\frac12r_++\frac32r_-.
\end{equation}
If the solution of Eqs.~(\ref{306.17}) is known, then the physical variables $\rho,u$ are given
by the expressions
\begin{equation}\label{310.1}
  \rho=\frac12(\sqrt{-r_+}\pm\sqrt{-r_-})^2,\qquad u=\pm 2\sqrt{r_+r_-},
\end{equation}
where both Riemann invariants are negative: $r_-\leq r_+\leq0$.

The Riemann invariants (\ref{306.15}) and the characteristic velocities (\ref{305.4}) are real
for $\rho\geq u$ ($\rho\geq0$ by definition), that is the inequalities $\rho\geq0, \rho\geq u$
define the {\it hyperbolicity domain} in the plane $(u,\rho)$ of physical variables.
Besides that, it is extremely important that the Riemann invariant $r_+$ reaches its maximal value $r_+=0$
along the $\rho$-axis where $u=0$. It means that its dependence on the physical variables is not
monotonous. We say that the $\rho$-axis $u=0$ cuts the hyperbolicity domain into two
 monotonicity regions $u<0$ and $u>0$. Correspondingly, the dependence of the physical variables on the Riemann
invariants is not single-valued---it is two-valued in our case of a single maximum of $r_+$,
if the solution of our hydrodynamics equations crosses the axis $u=0$.
As we shall see, this leads to important consequences in classification of wave structures evolving from
initial discontinuities.

%Solutions of these equations describe, in particular, rarefaction waves which will be discussed later.
Now we turn to derivation of the evolution equations for weakly nonlinear waves with small dispersion.

\section{Weakly nonlinear waves with small dispersion}\label{sec4}

The linear modes correspond to flows with fixed relationship between $\rho'$ and $u'$ and generalizations of these
waves to the nonlinear regime are simple waves with one of the Riemann invariants $r_{\pm}$ constant.
In the leading order, when the nonlinear and dispersive corrections are accounted in their main approximations,
we can add their effects in the resulting evolution equations. The small dispersive effects are described
by the last terms in the dispersion laws (\ref{305.11}) that can be transformed to the differential equations
for $\rho'$ by the replacements $\omega\to \rmi\prt_t$, $k\to-\rmi\prt_x$:
\begin{equation}\label{305.11b}
\rho'_t + \left(u_0-2\rho_0\pm\sqrt{\rho_0(\rho_0-u_0)}\right)\rho'_x
\mp \frac{1}{8\sqrt{\rho_0(\rho_0-u_0)}}\rho'_{xxx}=0.
\end{equation}
Therefore it is enough to consider now the weak nonlinear effects neglecting the dispersion. To simplify the notation,
we shall consider waves propagating along a uniform quiescent background with $\rho=\rho_0$, $u=u_0=0$.

\subsection{Kortweg-de Vries mode}

At first we shall consider waves with $r_+=\mathrm{const}$, and it is easy to find that far
enough from a localized wave pulse this Riemann invariant vanishes and the identity $r_-=0$ is fulfilled with
the accuracy up to the first order of small quantities $\rho'$ and $u$. Consequently, the equation for
$r_+$ is already satisfied with this accuracy and for the waves of density $\rho'$ we can substitute
$u=0$ into dispersionless expressions (\ref{305.4}) and (\ref{306.15}) for $r_-$ and $v_-$,
correspondingly, to find
$$
r_-\approx -2(\rho_0+\rho'),\qquad v_-\approx-3(\rho_0+\rho').
$$
Thus, dispersionless Hopf equation for this mode obtained from (\ref{306.17}) reads
$$
\rho'_t-3(\rho_0+\rho')\rho'_x=0,
$$
and addition of dispersion term from (\ref{305.11b}) for lower sign yields the
KdV equation
\begin{equation}\label{306.19}
  \rho'_t-3(\rho_0+\rho')\rho'_x-\frac{1}{8\rho_0}\rho'_{xxx}=0.
\end{equation}
Solution of the Riemann problem for this equation has very simple Gurevich-Pitaevskii type \cite{gp-73}.

\subsection{Modified Korteweg-de Vries mode}

In the mode with $r_-=-2\rho_0=\mathrm{const}$ we have to make calculations with accuracy
up to the second order with respect to $\rho'$. The condition $r_-=-2\rho_0$ gives us the
relationship
$$
u\approx 2\rho'-\frac{\rho'^{ 2}}{\rho_0},
$$
and its substitution into expressions (\ref{305.4}) and (\ref{306.15}) for $r_+$ and $v_+$
yields with the same accuracy
$$
r_+\approx-\frac{\rho'^{ 2}}{2\rho_0},\qquad v_+\approx -\rho_0-\frac{\rho'^{ 2}}{\rho_0}.
$$
Hence Eq.~(\ref{306.17}) for $r_+$ reduses to the dispersionless equation for the density
$$
\rho'_t-\left(\rho_0+\frac{\rho'^{ 2}}{\rho_0}\right)\rho'_x=0,
$$
and addition of dispersion term from (\ref{305.11b}) for upper sign yields the
mKdV equation
\begin{equation}\label{306.19}
  \rho'_t--\left(\rho_0+\frac{\rho'^{ 2}}{\rho_0}\right)\rho'_x+\frac{1}{8\rho_0}\rho'_{xxx}=0.
\end{equation}
For this mode the solution of the Riemann problem \cite{kamch-12,ehs-17} is much more
complicated and this fact suggests that the Riemann problem for the DNLS equation
must differ considerably from that for the NLS equation \cite{eggk-95}. To find this solution,
we have to obtain the periodic solutions in convenient for us form parameterized by the
Riemann invariants of the Whitham modulation equations and to derive these modulation equations.
Actually, that was done in Refs.~\cite{kamch-90a,kamch-90b}, however, for completeness we shall
reproduce here briefly these results with some improvements.

\section{Periodic solutions of the DNLS equation}\label{sec5}

The finite-gap integration method (see, e.g., \cite{kamch-2000}) of finding periodic solutions
is based on possibility of representing the DNLS equation (\ref{DNLS}) as a compatibility condition of two
systems of linear equations with a spectral parameter $\la$
\begin{equation}\label{lax1}
    \frac{\partial}{\partial x}
    \left(
    \begin{array}{c}
        {\psi}_1 \\
        {\psi}_2 \\
    \end{array}
    \right)
    =\mathbb{U}
    \left(
    \begin{array}{c}
        {\psi}_1 \\
        {\psi}_2 \\
    \end{array}
    \right), \quad
    \frac{\partial}{\partial t}
    \left(
    \begin{array}{c}
        {\psi}_1 \\
        {\psi}_2 \\
    \end{array}
    \right)
    =\mathbb{V}
    \left(
    \begin{array}{c}
        {\psi}_1 \\
        {\psi}_2 \\
    \end{array}
    \right),
\end{equation}
where
\begin{equation}
        \mathbb{U=}\left(
    \begin{array}{cc}
        {F} & {G} \\
        {H} & -{F} \\
    \end{array}
    \right)\;,\quad
    \mathbb{V=}\left(
    \begin{array}{cc}
        {A} & {B} \\
        {C} & -{A} \\
    \end{array}
    \right),
\end{equation}
with
\begin{equation}\label{307.5}
     \eqalign{
     {F}=-2{\rm i}\lambda^2,\quad {G}=2\lambda \Psi,\quad {H}=2\lambda \Psi^*, \\
     {A}=-{\rm i}\left(4\lambda^4+2\lambda^2|\Psi|^2\right), \quad
     {B}=4\lambda^3\Psi+\lambda\left({\rm i} \Psi_x+2|\Psi|^2\Psi\right), \\
     {C}=4\lambda^3\Psi^*-\lambda\left({\rm i} \Psi^*_x-2|\Psi|^2\Psi^*\right).
    }
\end{equation}
The compatibility condition of linear systems (\ref{lax1}),
\begin{equation}\label{307.8}
    \mathbb{U}_t-\mathbb{V}_x+[\mathbb{U},\mathbb{V}]=0,
\end{equation}
where $[\cdot,\cdot]$ is a commutator of matrices, is equivalent to the DNLS equation.

If we denote as $(\psi_1,\psi_2)$ and $(\bar{\psi_1},\bar{\psi}_2)$ two basis solutions of
linear systems (\ref{lax1}) and introduce a matrix of `squared basis functions'
\begin{equation}\label{307.6a}
  \mathbb{W}=\left(
               \begin{array}{cc}
                 -\rmi f & h \\
                 h & \rmi f \\
               \end{array}
             \right),
\end{equation}
where
\begin{equation}\label{307.6b}
  f=-\frac{\rmi}2(\psi_1\bar{\psi}_2+\psi_2\bar{\psi}_1),\quad g=\psi_1\bar{\psi}_1,\quad h=-\psi_2\bar{\psi}_2,
\end{equation}
then equations for these functions can be written in matrix form
\begin{equation}\label{307.7}
    \mathbb{W}_x=[\mathbb{U},\mathbb{W}],\qquad \mathbb{W}_t=[\mathbb{V},\mathbb{W}].
\end{equation}
It is known that the characteristic polynomial
\begin{equation}\label{307.9}
  \mathrm{det}(iw\cdot\mathbb{I}-\mathbb{W})=-w^2+f^2-gh
\end{equation}
does not depend on $t$ and $x$ (in our simple case it can be checked by a simple calculation and the general
proof of this theorem can be found, e.g., in appendix B of Ref.~\cite{kamch-2014}). Hence, it defines the curve
\begin{equation}\label{307.10}
  w^2=P(\la),\qquad P(\la)=f^2-gh,
\end{equation}
where $P(\la)$ depends on $\la$ only.

Periodic solutions are distinguished by the condition that $P(\la)$ be a polynomial in $\la$, and then the structure of
the matrix elements (\ref{307.5}) suggests that $f,g,h$ must also be polynomials in $\la$. The simplest
one-phase solution corresponds to the polynomials $f,g,h$ in the form
\begin{equation}\label{307.11}
  f=\la^4-f_1\la^2+f_2,\quad g=\la(\la^2-\mu/2)\Psi,\quad h=\la(\la^2-\mu^*/2)\Psi^*.
\end{equation}
The functions ${f}_1(x,t)$, ${f}_2(x,t)$, $\mu(x,t)$ and $\mu^*(x,t)$
are unknown yet, but we shall see soon that  $\mu(x,t)$ and $\mu^*(x,t)$ are complex conjugate,
whence the notation. Then the polynomial $P(\la)$ can be written as
\begin{equation}\label{307.12}
    P(\lambda) = \prod_{i=1}^4\left(\lambda^2-\lambda_i^2\right)
    = \lambda^8-s_1 \lambda^6+s_2 \lambda^4-s_3\lambda^2+s_4,
\end{equation}
where $s_i$ are symmetric functions of the four
zeroes $\lambda_i^2$ of the polynomial,
\begin{equation}\label{}
         s_1=\sum_i\lambda_i^2,\quad s_2=\sum_{i<j}\lambda_i^2\lambda_j^2,
    \quad s_3=\sum_{i<j<k}\lambda_i^2\lambda_j^2\lambda_k^2, \quad
     s_4=\lambda_1^2\lambda_2^2\lambda_3^2\lambda_4^2,
\end{equation}
and the identity (\ref{307.10}) yields the conservation laws
\begin{equation}\label{307.13}
  \eqalign{
  s_1=2f_1+\nu,\quad
  s_2=f_1^2+sf_2+\nu(\mu+\mu^*),\\
  s_3=2f_1f_2-\nu\mu\mu^*, \quad
  s_4=f_2^2,
  }
\end{equation}
where $\nu=|\Psi|^2$. This system permits one to express ${f}_1(x,t)$, ${f}_2(x,t)$,
$\mu(x,t)$ and $\mu^*(x,t)$ as functions of $\nu$:
\begin{eqnarray}\label{308.1}
f_1=(s_1-\nu)/2,\qquad f_2=\pm\sqrt{s_4},\\
\label{308.2}
\mu,\mu^*=\frac1{8\nu}\left(4s_2\pm8\sqrt{s_4}-(\nu-s_1)^2+i\sqrt{-\mathcal{R}(\nu)}\right),
\end{eqnarray}
where the polynomial
\begin{equation}\label{308.3}
\eqalign{
  \mathcal{R}(\nu)=\nu^4-4s_1\nu^3+(6s_1^2-8s_2\pm48\sqrt{s_4})\nu^2\\
  -(4s_1^3-16s_1s_2+64s_3\pm32s_1\sqrt{s_4})\nu+(-s_1^2+4s_2\pm8\sqrt{s_4})^2
  }
\end{equation}
is called a resolvent of the polynomial $P(\la)$ since its zeroes $\nu_i$ are related with the zeroes
$\la_j$ of $P(\la)$ by symmetric formulae: the upper signs ($+$) in (\ref{308.3}) corresponds to the
zeroes
\begin{equation}\label{zeros1}
    \eqalign{
    \nu_1  =(-\lambda_1+\lambda_2+\lambda_3+\lambda_4)^2, \quad
    &\nu_2  =(\lambda_1-\lambda_2+\lambda_3+\lambda_4)^2, \\
    \nu_3  =(\lambda_1+\lambda_2-\lambda_3+\lambda_4)^2, \quad
    &\nu_4  =(\lambda_1+\lambda_2+\lambda_3-\lambda_4)^2,
    }
\end{equation}
and the lower signs ($-$) in equation (\ref{308.3}) correspond to the zeroes
\begin{equation}\label{zeros2}
    \eqalign{
    \nu_1  =(-\lambda_1+\lambda_2+\lambda_3-\lambda_4)^2, \quad
    &\nu_2  =(\lambda_1-\lambda_2+\lambda_3-\lambda_4)^2, \\
    \nu_3  =(\lambda_1+\lambda_2-\lambda_3-\lambda_4)^2, \quad
    &\nu_4  =(\lambda_1+\lambda_2+\lambda_3+\lambda_4)^2. \\
    }
\end{equation}
This can be proved by a simple check of the Vi\`ete formulae. In both cases the zeroes
are ordered according to $\nu_1\leq\nu_2\leq\nu_3\leq\nu_4$ for
$\la_1\leq\la_2\leq\la_3\leq\la_4\leq0$.

From the components
\begin{equation}\label{308.4}
g_x=2\rmi Gf+2Fg,\qquad g_t=2\rmi Bf+2Ag
\end{equation}
of the matrix equations (\ref{307.7}) at $\la=\mu^{1/2}$ we find that $\mu$
satisfies the equations
\begin{equation}\label{308.5}
  \mu_x=4\rmi\sqrt{P(\mu^{1/2})},\quad \mu_t=8\rmi(2f_1+\nu)\sqrt{P(\mu^{1/2})}=2s_1\mu_x,
\end{equation}
where we have used the first equation (\ref{307.13}). Consequently, $\mu$ depends on the
phase $\xi=x-Vt$ only, where $V=-2s_1=-2\sum\la_i^2$. Then the variable $\nu$ also depends
on $\xi$ only. Substitution of $g=\la(\la^2-\mu)\Psi$ into the first equation (\ref{308.4})
gives $\Psi_x=-4\rmi \Psi(f_1-\mu)$, so that $\nu_x=4\rmi\nu(\mu-\mu^*)$, and, with the use of
(\ref{308.2}), we obtain equation for $\nu$,
\begin{equation}\label{308.7}
  \frac{d\nu}{d\xi}=\sqrt{-\mathcal{R}(\nu)},\quad \xi=x-Vt,\quad V=-\sum\la_i^2=\frac14\sum\nu_i.
\end{equation}
The real solutions of this equation correspond to oscillations of $\nu$ within the intervals where
$-\mathcal{R}(\nu)\geq0$. We shall discuss two possibilities separately.

(A) At first we shall consider the periodic solution corresponding to oscillations
of $\nu$ in the interval
\begin{equation}\label{eq18}
    \nu_1\leq \nu \leq \nu_2.
\end{equation}
Standard calculation yields after some algebra the solution in terms of Jacobi
elliptic functions:
\begin{equation}\label{eq20}
    \nu=\nu_2-\frac{(\nu_2-\nu_1)\cn^2(\theta,m)}{1+\frac{\nu_2-\nu_1}{\nu_4-\nu_2}\sn^2(\theta,m)},
\end{equation}
where it is assumed that $\nu(0)=\nu_1$, and
\begin{equation}\label{eq21}
    \theta=\sqrt{(\nu_3-\nu_1)(\nu_4-\nu_2)}\,\xi/2,\qquad
    m=\frac{(\nu_4-\nu_3)(\nu_2-\nu_1)}{(\nu_4-\nu_2)(\nu_3-\nu_1)},
\end{equation}
%\begin{equation}\label{eq22}
%    m=\frac{(\nu_4-\nu_3)(\nu_2-\nu_1)}{(\nu_4-\nu_2)(\nu_3-\nu_1)},
%\end{equation}
$\cn$ and $\sn$ being Jacobi elliptic functions.
The wavelength of the oscillating function (\ref{eq20}) is
\begin{equation}\label{eq23}
    L=\frac{4K(m)}{\sqrt{(\nu_3-\nu_1)(\nu_4-\nu_2)}}=\frac{K(m)}{\sqrt{(\la_3^2-\la_1^2)(\la_4^2-\la_2^2)}},
\end{equation}
where $K(m)$ is the complete elliptic integral of the first kind.

In the limit $\nu_3\to \nu_2$ ($m\to1$) the wavelength
tends to infinity and the solution (\ref{eq20}) acquires the soliton form
\begin{equation}\label{eq24}
    \nu=\nu_2-\frac{\nu_2-\nu_1}{\cosh^2\theta+\frac{\nu_2-\nu_1}{\nu_4-\nu_2}\sinh^2\theta}.
\end{equation}
This is a ``dark soliton'' for the variable $\nu$.

The limit $m\to0$ can be reached in two ways.

(i) If $\nu_2\to \nu_1$, then the solution transforms into a linear harmonic
wave
\begin{equation}\label{eq25}
    \nu\cong \nu_2-\frac12(\nu_2-\nu_1)\cos(k\xi),\quad
    k =\sqrt{(\nu_3-\nu_1)(\nu_4-\nu_1)}.
\end{equation}

(ii) If $\nu_4=\nu_3$ but $\nu_1\neq \nu_2$, then  then we arrive at
the nonlinear trigonometric solution:
\begin{equation}\label{eq26}
    \nu=\nu_2-\frac{(\nu_2-\nu_1)\cos^2\theta}{1+\frac{\nu_2-\nu_1}{\nu_3-\nu_2}\sin^2\theta},\quad
    \theta=\sqrt{(\nu_3-\nu_1)(\nu_3-\nu_2)}\,\xi/2.
\end{equation}
If we take the limit $\nu_2-\nu_1\ll \nu_3-\nu_1$ in this solution, then we
return to the small-amplitude limit (\ref{eq25}) with $\nu_4=\nu_3$. On
the other hand, if we take here the limit $\nu_2\to \nu_3=\nu_4$, then the
argument of the trigonometric functions becomes small and we can
approximate them by the first terms of their series expansions. This
corresponds to an algebraic soliton of the form
\begin{equation}\label{eq27}
   \nu=\nu_2-\frac{\nu_2-\nu_1}{1+(\nu_2-\nu_1)^2\xi^2/4}.
\end{equation}

(B) In the second case, the variable $\nu$ oscillates in the interval
\begin{equation}\label{eq28}
    \nu_3\leq \nu\leq \nu_4\; .
\end{equation}
Here again, a standard calculation yields
\begin{equation}\label{eq30}
   \nu=\nu_3+\frac{(\nu_4-\nu_3)\cn^2(\theta,m)}{1+\frac{\nu_4-\nu_3}{\nu_3-\nu_1}\sn^2(\theta,m)}
\end{equation}
with the same definitions (\ref{eq21}) and (\ref{eq23})
of $\theta$, $m$, and $L$. In this case we have $\nu(0)=\nu_4$.
In the soliton limit $\nu_3\to \nu_2$ ($m\to1$) we get
\begin{equation}\label{eq31}
   \nu=\nu_2+\frac{\nu_4-\nu_2}{\cosh^2\theta+\frac{\nu_4-\nu_2}{\nu_2-\nu_1}\sinh^2\theta}.
\end{equation}
This is a ``bright soliton'' for the variable $\nu$.

Again, the limit $m\to0$ can be reached in two ways.

(i) If $\nu_4\to \nu_3$, then we obtain a small-amplitude harmonic wave
\begin{equation}\label{eq32}
    \nu \cong \nu_3+\frac12(\nu_4-\nu_3)\cos(k\xi),\quad
    k=\sqrt{(\nu_3-\nu_1)(\nu_3-\nu_2)}.
\end{equation}

(ii) If $\nu_2=\nu_1$, then we obtain another nonlinear trigonometric solution,
\begin{equation}\label{eq33}
    \nu=\nu_3+\frac{(\nu_4-\nu_3)\cos^2\theta}{1+\frac{\nu_4-\nu_3}{\nu_3-\nu_1}\sin^2\theta},\quad
    \theta=\sqrt{(\nu_3-\nu_1)(\nu_4-\nu_1)}\,\xi/2.
\end{equation}
If we assume that $\nu_4-\nu_3\ll \nu_4-\nu_1$, then we reproduce
the small-amplitude limit (\ref{eq32}) with $\nu_2=\nu_1$. On the other hand,
in the limit $\nu_3\to \nu_2=\nu_1$ we obtain the algebraic soliton solution:
\begin{equation}\label{eq34}
    \nu=\nu_1+\frac{\nu_4-\nu_1}{1+(\nu_4-\nu_1)^2\xi^2/4}.
\end{equation}

For both cases (\ref{zeros1}), (\ref{zeros2})
we have the identities
\begin{equation}\label{}
    m=\frac{(\nu_4-\nu_3)(\nu_2-\nu_1)}{(\nu_4-\nu_2)(\nu_3-\nu_1)}
        =\frac{(\lambda_4^2-\lambda_3^2)(\lambda_2^2-\lambda_1^2)}{(\lambda_4^2-\lambda_2^2)(\lambda_3^2-\lambda_1^2)}.
\end{equation}

The importance of this form of periodic solutions of our equation is related with the fact that
the parameters $\lambda_j$, connected with $\nu_i$ by the formulae (\ref{zeros1}), (\ref{zeros2}),
can play the role of Riemann invariants in the Whitham theory of modulations.

\section{Whitham modulation equations}\label{sec6}

In modulated waves the parameters $\la_i$ become slowly varying functions of
the space and time variables and their evolution is governed by the
Whitham modulation equations. Whitham showed in
Ref.~\cite{whitham-65} that these equations can be obtained
by averaging the conservation laws of the full nonlinear system over
fast oscillations (whose wavelength $L$ changes slowly along the total
wave pattern).  Generally speaking, in cases where the periodic
solution is characterized by four parameters, this averaging procedure
leads to a system of four equations of the type
$\nu_{i,t}+\sum_jv_{ij}(\nu_1,\nu_2,\nu_3,\nu_4)\nu_{j,x}=0$ with 16 entries in
the ``velocity matrix'' $v_{ij}$.  However, for the case of completely
integrable DNLS equation, this system of four equations
reduces to a diagonal {\it Riemann form} for the {\it Riemann invariants}
$\la_i$'s, similar to what occurs for the usual Riemann invariants
of non-dispersive waves (see Eqs.~(\ref{306.17})).
We shall derive the modulation Whitham equations by the method developed
in Refs.~\cite{kamch-90b,kamch-2000}.

First of all, we notice that with the use of (\ref{307.8}) and (\ref{308.4}) it
is easy to prove the identity
\begin{equation}\label{275.47}
  \frac{\prt}{\prt t}\left(\sqrt{P(\la)}\cdot\frac{G(\la)}{g(\la)}\right)-
  \frac{\prt}{\prt x}\left(\sqrt{P(\la)}\cdot\frac{B(\la)}{g(\la)}\right)=0,
\end{equation}
where we have introduced under the derivative signs the constant on periodic solutions
factor $\sqrt{P(\la)}$ to transform the identity (\ref{307.10}) to the
form
$$
  \left(\frac{f}{\sqrt{P(\la)}}\right)^2-
\frac{g}{\sqrt{P(\la)}}\cdot\frac{h}{\sqrt{P(\la)}}=1,
$$
so that the right-hand side is independent of the variations of
$\la_i$ in a modulated wave, hence the densities and fluxes in the
conservation laws can change due to modulations only, as it should be,
and any changes caused by $\lambda$-dependent normalization of the
$f,g,h$-functions are excluded.
We shall use the equation (\ref{275.47}) as the generating function of the
conservation laws of the DNLS equation: a series expansion
in inverse powers of $\la$ gives an infinite number of conservation
laws of this completely integrable system.

Substitution of Eqs.~(\ref{307.8}) and (\ref{308.4})
into (\ref{275.47}) and its simple transformation gives
\begin{equation}\nonumber
  \frac{\prt}{\prt t}\left(\frac{\sqrt{P(\la)}}{\la^2-\mu}\right)-
  \frac{\prt}{\prt x}
\left[\sqrt{P(\la)}\left(2+\frac{s_1}{\la^2-\mu}\right)\right]=0,
\end{equation}
Averaging of the density and of the flux in this expression over one
wavelength $L$
\begin{equation}\label{L}
  L=\oint\frac{\rmd \mu}{4\sqrt{-P(\mu^{1/2})}}
\end{equation}
according to the rule
$$
\langle\left\{\ldots\right\}\rangle=\int_0^L\left\{\ldots\right\}\frac{\rmd x}L=
\frac1L\oint\left\{\ldots\right\}\frac{\rmd x}{\rmd\mu}\rmd\mu=
\frac1L\oint\left\{\ldots\right\}\frac{\rmd \mu}{4\sqrt{-P(\mu^{1/2})}}
$$
yields the generating function of the
{\it averaged} conservation laws:
\begin{equation}\label{276.48}
\eqalign{
  \frac{\prt}{\prt t}
\left[\frac{\sqrt{P(\la)}}L\oint\frac{\rmd\mu}{4(\la^2-\mu)\sqrt{-P(\mu^{1/2})}}\right]\\
  -
  \frac{\prt}{\prt x}
\left[\sqrt{P(\la)}\left(2+\frac{s_1}{L}\oint\frac{\rmd\mu}{4(\la^2-\mu)\sqrt{-P(\mu^{1/2})}} \right)
  \right]=0.
  }
\end{equation}
The condition that in the limit $\la\to\la_i$ the singular terms
cancel yields
\begin{equation}\label{equa65}
\eqalign{
  \oint\frac{d\mu}{4(\la^2_i-\mu)\sqrt{-P(\mu^{1/2})}}\cdot\frac{\prt\la^2_i}{\prt t}\\
  - \left(2L+{s_1}\oint\frac{d\mu}{4(\la_i-\mu)\sqrt{-P(\mu^{1/2})}}\right)
\cdot\frac{\prt\la^2_i}{\prt x}=0.
}
\end{equation}
From the definition (\ref{L}) of $L$ one obtains
$$
\oint \frac{d\mu}{4(\la^2_i-\mu)\sqrt{-P(\mu^{1/2})}}=-2\, \frac{\prt L}{\prt\la^2_i},
$$
which makes it possible to cast Eq.~(\ref{equa65}) in the form of a
Whitham equation for the variables $\la_i$:
\begin{equation}\label{276.56}
  \frac{\prt\la_i}{\prt t}+v_i\frac{\prt\la_i}{\prt x}=0,
\end{equation}
where the Whitham velocities $v_i$ are given by
\begin{equation}\label{276.57}
  v_i=-{s_1}+\frac{L}{\prt L/\prt\la^2_i},\quad \mbox{for}\quad
i=1,2,3,4.
\end{equation}

The values $\la_i$ of the spectral parameters are well-defined Riemann invariants of the
Whitham system of modulation equations, however, they do not suit well enough to the
problems with matching of modulated cnoidal waves and smooth dispersionless solutions.
Therefore it is more convenient to define new set of Whitham invariants by using simple
fact that any function of a single argument $\la_i$ is also a Riemann invariant. We define
the new Riemann invariants by the formulae
\begin{equation}\label{312.1}
  r_i=-2\la_i^2,\quad i=1,2,3,4.
\end{equation}
They are negative and ordered according to $r_1\leq r_2\leq r_3\leq r_4\leq0$ for
$\la_1\leq\la_2\leq\la_3\leq\la_4\leq0$. The parameters $\nu_i$ of the periodic solutions
of the DNLS equation are expressed in terms of $r_i$ as
\begin{equation}\label{312.3a}
    \eqalign{
    \nu_1  =(-\sqrt{-r_1}+\sqrt{-r_2}+\sqrt{-r_3}+\sqrt{-r_4})^2/2, \\
    \nu_2  =(\sqrt{-r_1}-\sqrt{-r_2}+\sqrt{-r_3}+\sqrt{-r_4})^2/2, \\
    \nu_3  =(\sqrt{-r_1}+\sqrt{-r_2}-\sqrt{-r_3}+\sqrt{-r_4})^2/2, \\
    \nu_4  =(\sqrt{-r_1}+\sqrt{-r_2}+\sqrt{-r_3}-\sqrt{-r_4})^2/2,
    }
\end{equation}
or
\begin{equation}\label{312.3b}
    \eqalign{
    \nu_1  =(-\sqrt{-r_1}+\sqrt{-r_2}+\sqrt{-r_3}-\sqrt{-r_4})^2/2, \\
    \nu_2  =(\sqrt{-r_1}-\sqrt{-r_2}+\sqrt{-r_3}-\sqrt{-r_4})^2/2, \\
    \nu_3  =(\sqrt{-r_1}+\sqrt{-r_2}-\sqrt{-r_3}-\sqrt{-r_4})^2/2, \\
    \nu_4  =(\sqrt{-r_1}+\sqrt{-r_2}+\sqrt{-r_3}+\sqrt{-r_4})^2/2. \\
    }
\end{equation}
The phase velocity and the wavelength are given by
\begin{equation}\label{312.2}
  V=\frac12\sum_{i=1}^4r_i,\quad L=\frac{2K(m)}{\sqrt{(r_4-r_2)(r_3-r_1)}},\quad
  m=\frac{(r_4-r_3)(r_2-r_1)}{(r_4-r_2)(r_3-r_1)}.
\end{equation}
The Whitham modulation equations read
\begin{equation}\label{312.7}
  \frac{\prt r_i}{\prt t}+v_i\frac{\prt r_i}{\prt x}=0,\quad i=1,2,3,4,
\end{equation}
where the Whitham velocities $v_i$ are given by
\begin{equation}\label{276.57}
  v_i=-{s_1}-\frac{L}{2\prt L/\prt r_i},\quad \mbox{for}\quad
i=1,2,3,4,
\end{equation}
and substitution of $L$ from (\ref{312.2}) gives after simple calculation
the following explicit expressions
\begin{equation}\label{312.8}
\eqalign{
     v_1=\frac12\sum_{i=1}^{4}r_i
-\frac{(r_4-r_1)(r_2-r_1)
K(m)}{(r_4-r_1)K(m) -(r_4-r_2)E(m)}, \\
     v_2=\frac12\sum_{i=1}^{4}r_i
+\frac{(r_3-r_2)(r_2-r_1)K(m)}
{(r_3-r_2) K(m)-(r_3-r_1)E(m)}, \\
     v_3=\frac12\sum_{i=1}^{4}r_i-
\frac{(r_4-r_3)(r_3-r_2)K(m)}
{(r_3-r_2)K(m)- (r_4-r_2)E(m)}, \\
     v_4=\frac12\sum_{i=1}^{4}r_i
+\frac{(r_4-r_3)(r_4-r_1)K(m)}
{(r_4-r_1)K(m) -(r_3-r_1)E(m)},
}
\end{equation}
where $K(m)$ and $E(m)$ are complete elliptic integrals of the first
and second type, respectively.

In a modulated wave representing a dispersive shock wave, the Riemann invariants change
slowly with $x$ and $t$. The dispersive shock wave occupies a space interval at whose edges
two of the Riemann invariants are equal to each other.
The soliton edge corresponds to $r_3=r_2$ $(m=1)$ and at this edge
the Whitham velocities are given by
\begin{equation}\label{sol-limit}
\eqalign{
    {v_1}  =\frac12(3r_1+r_4), \quad
   {v_4}  =\frac12(r_1+3r_4),\\
    {v_2}  ={v_3}=\frac12(r_1+2r_2+r_4).
    }
\end{equation}
The opposite limit $m=0$ can be obtained in two ways. If
$r_3=r_4$, then we get
\begin{equation}\label{314.3}
\eqalign{
    {v_1} =\frac12(3r_1+r_2), \quad {v_2}=\frac12(r_1+3r_2), \\
    {v_3} ={v_4}=2r_4+\frac{(r_2-r_1)^2}{2(r_1+r_2-2r_4)},
    }
\end{equation}
and if $r_2=r_1$, then
\begin{equation}\label{314.2}
\eqalign{
    {v_1}  ={v_2}=2r_1+\frac{(r_4-r)^2}{2(r_3+r_4-2r_1)}, \\
    {v_3}  =\frac12(3r_3+r_4), \quad {v_4}=\frac12(r_3+3r_4).
    }
\end{equation}
From these equations it is clear that at the edges of the oscillatory zone
the Whitham equation for two Riemann invariants coincide with those for the
dispersionless equations, that is the oscillatory zone can match at its edges
with smooth solutions of the dispersionless equations.

Now we are ready to discuss the key elements from which consists any wave
structure evolving from an initial discontinuity.

\section{Elementary wave structures}\label{sec7}

Our aim in this paper is to develop the method of derivation of the asymptotic solution
of the  DNLS evolution problem for a discontinuous step-like initial conditions
\begin{equation}\label{init}
\rho(x,0)=\cases{\rho^L &for $x < 0$\\
\rho^R&for $x>0$\\} \qquad
u(x,0)=\cases{u^L &for $x < 0$\\
u^R&for $x>0$\\}.
\end{equation}
As we shall see, evolution of this step-like pulse leads to formation of quite
complex wave structures consisting of several simpler elements of simple wave type
with only one Riemann invariant changing. Therefore we shall
first describe these elements in the present section.

\subsection{Rarefaction waves}

For smooth enough dependence of wave parameters on $x$ and $t$, we can neglect the
dispersion effects and use the dispersionless equations derived in section~\ref{sec3}.
First of all, we notice that the system (\ref{306.17}) has a trivial solution for which
$r_+=\mathrm{const}$ and $r_-=\mathrm{const}$. We shall call such a solution
a ``plateau'' because it corresponds to a uniform flow with constant density and flow
velocity given by (\ref{310.1}).

The initial conditions (\ref{init}) do not contain any parameters with dimension of time or length.
Therefore solutions of equations (\ref{306.17}) can depend on the self-similar variable
$\zeta=x/t$ only, that is $r_{\pm}=r_{\pm}(\zeta)$, and then this system reduces to
\begin{equation}\label{310.2}
    \left({v_+}-\zeta\right)\frac{dr_+}{d\zeta}=0, \quad \left({v_-}-\zeta\right)\frac{dr_-}{d\zeta}=0.
\end{equation}
Evidently, these equations have solutions with one of the Riemann invariants constant and the other
one changes in such a way, that the corresponding velocity equals to $\zeta=x/t$. To be definite,
let us consider the solution
\begin{equation}\label{310.5}
  r_+=r_+^0=\mathrm{const}, \qquad v_-=\frac12r_+^0+\frac32r_-=\zeta=\frac{x}t.
\end{equation}
Consequently, $r_-$ depends on $x/t$ as
\begin{equation}\label{310.6}
  r_-(x,t)=-\frac13r_+^0+\frac23\cdot\frac{x}t,
\end{equation}
and according to Eqs.~(\ref{310.1}) the physical variables are given by
\begin{equation}\label{310.7}
\eqalign{
  \rho_{\pm}(x,t)=\frac12\left(\sqrt{-r_+^0}\pm\sqrt{\frac13r_+^0-\frac23\cdot\frac{x}t}\right)^2,\\
  u_{\pm}(x,t)=\pm2\sqrt{-r_+^0\left(\frac13r_+^0-\frac23\cdot\frac{x}t\right)}.
  }
\end{equation}
Here the single solution (\ref{310.5}) of equations written in Riemann form yields two
solutions (\ref{310.7}) in physical variables which we distinguish by the indices $\pm$.
These rarefaction waves match to the plateau solutions at their left and right edges. At
both edges the invariant $r_+$ has the same value $r_+=r_+^0$ whereas we have $r_-=r_-^L$ at the
left boundary and $r_-=r_-^R$ at the right boundary. Correspondingly, the above two
solutions match to the values of the density
\begin{equation}\label{310.9}
  \eqalign{
  (i) \quad \rho_+^L=\frac12\left(\sqrt{-r_+^0}+\sqrt{-r_-^L}\right)^2,\quad
 \rho_+^R=\frac12\left(\sqrt{-r_+^0}+\sqrt{-r_-^R}\right)^2,\\
  (ii) \quad \rho_-^L=\frac12\left(\sqrt{-r_+^0}-\sqrt{-r_-^L}\right)^2,\quad
  \rho_-^R=\frac12\left(\sqrt{-r_+^0}-\sqrt{-r_-^R}\right)^2,
  }
\end{equation}
and similar formulae can be written for the flow velocities $u_{\pm}^{L,R}$.
The edge points propagate with velocities
\begin{equation}\label{310.10}
  s^L=v_-(r_+^0,r_-^L)=\frac12r_+^0+\frac32r_-^L,\quad
  s^R=v_-(r_+^0,r_-^R)=\frac12r_+^0+\frac32r_-^R.
\end{equation}
Since $r_-^R<r_+^R=r_+^0$, we always have $s^R<r_+^0$.

\begin{figure}[t] \centering
\includegraphics[width=10cm]{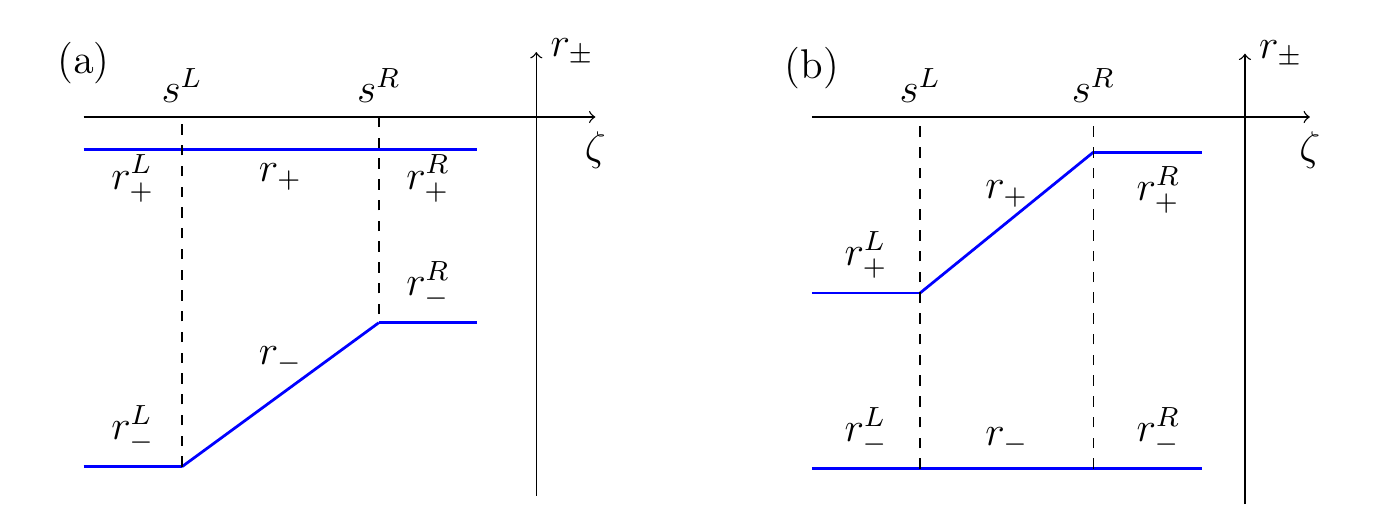}
\caption{Diagrams of Riemann invariants for the rarefaction wave solutions of the DNLS equation
in the dispersionless limit: (a) $r_+=\mathrm{const}$; (b) $r_-=\mathrm{const}$.}
\label{Fig1}
\end{figure}

In a similar way we obtain the second solution
\begin{equation}\label{310.12}
  r_+=-\frac13r_-^0+\frac23\cdot\frac{x}t,\quad r_-=r_-^0=\mathrm{const},
\end{equation}
hence
\begin{equation}\label{311.1}
\eqalign{
  \rho_{\pm}(x,t)=\frac12\left(\sqrt{\frac13r_-^0-\frac23\cdot\frac{x}t}\pm \sqrt{-r_-^0}\right)^2,\\
  u_{\pm}(x,t)=\pm2\sqrt{-r_-^0\left(\frac13r_-^0-\frac23\cdot\frac{x}t\right)}.
  }
\end{equation}
In this case we have
\begin{equation}\label{311.2}
  r_-^0\leq s^L=\frac32r_+^L+\frac12r_-^0< s^R=\frac32r_+^R+\frac12r_-^0<\frac12r_-^0.
\end{equation}

\begin{figure}[t] \centering
\includegraphics[width=10cm]{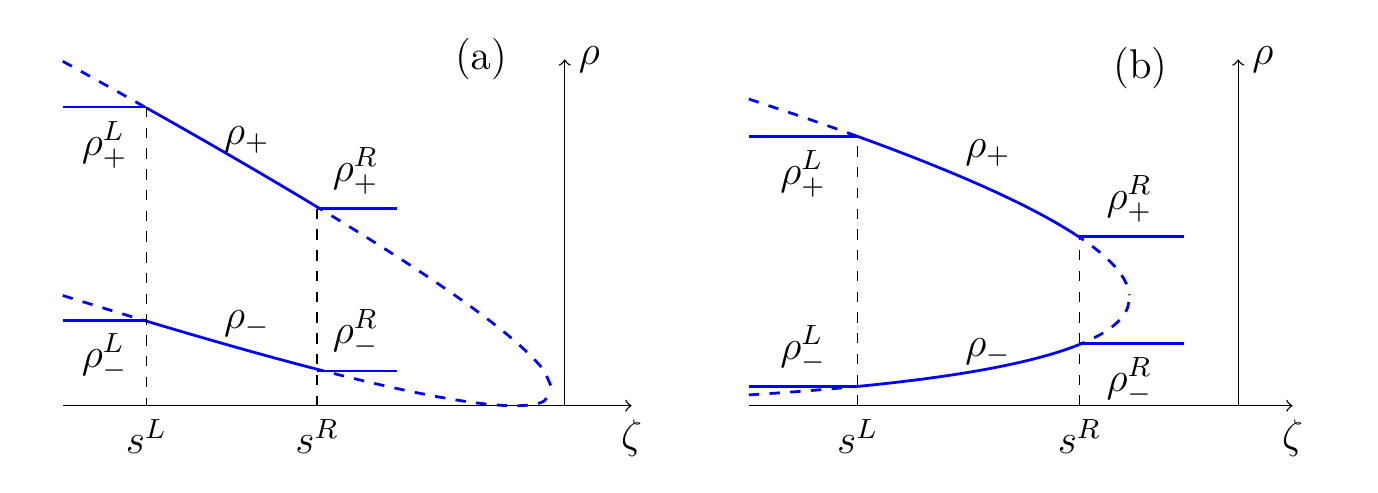}
\caption{Density distributions in the rarefaction wave solutions of the DNLS equation: (a) $r_+=\mathrm{const}$; (b) $r_-=\mathrm{const}$.
}
\label{Fig2}
\end{figure}

Diagrams of the Riemann invariants
for these rarefaction wave solutions are shown in Fig.~\ref{Fig1}: the case (a)
corresponds to Eqs.~(\ref{310.5}), (\ref{310.6}) and the case (b) to Eqs.~(\ref{310.12}).
Corresponding plots of densities are demonstrated in Fig.~\ref{Fig2} by thick lines
together with plateau distributions at the edges of the rarefaction waves. Dashed thick lines
show both branches of the solutions (\ref{310.7}) and (\ref{311.1}). It is worth noticing that
the edge velocities are determined by the Riemann invariants only and do not depend on the
choice of the branch into which the Riemann invariants are mapped.

\begin{figure}[t] \centering
\includegraphics[width=10cm]{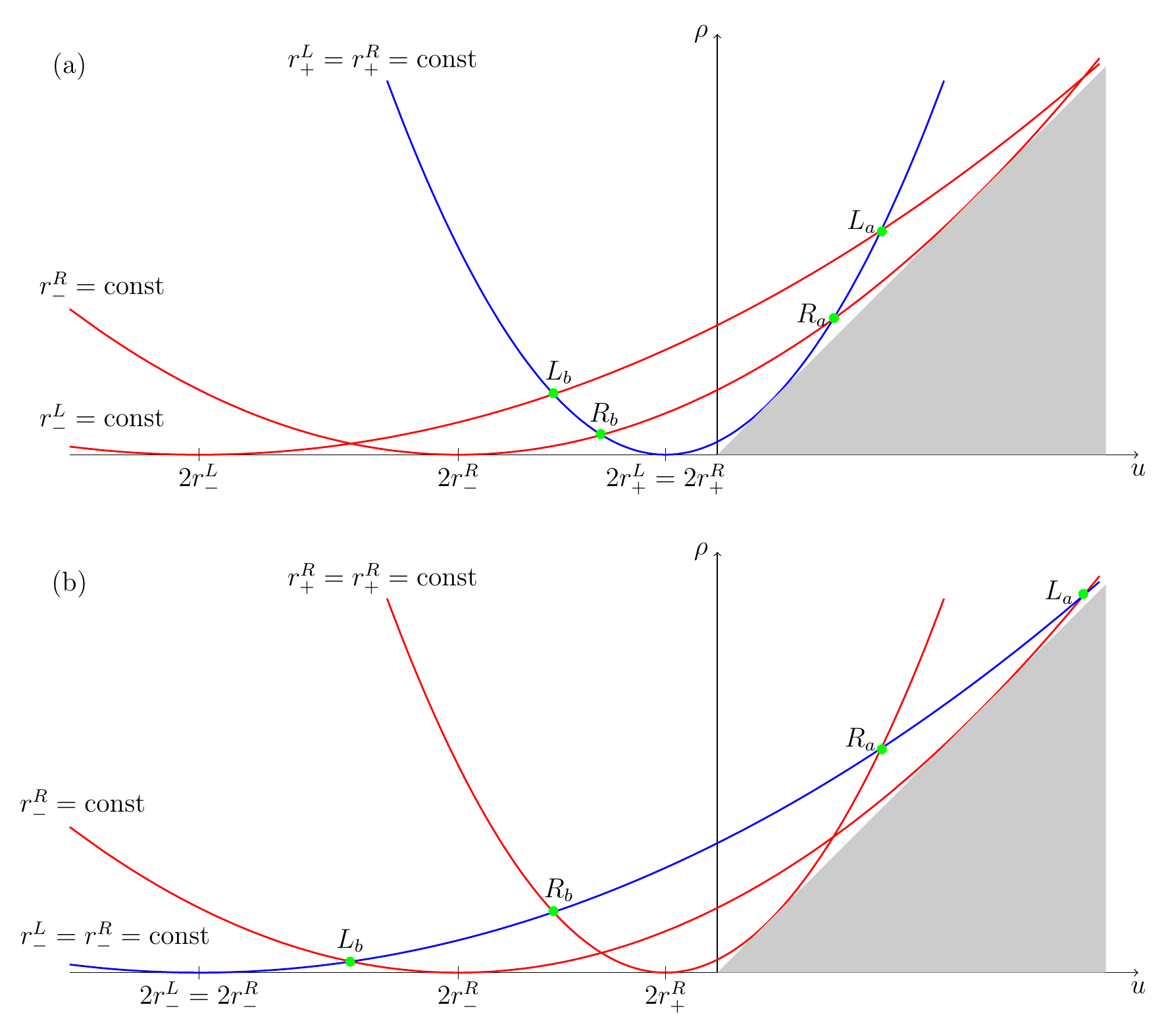}
\caption{Curves of constant Riemann invariants in the $(u,\rho)$-plane and transitions corresponding to
rarefaction waves. Plot (a) represents the rarefaction waves with $r_+=\mathrm{const}$
and (b) with $r_-=\mathrm{const}$. Grey areas $u>\rho$ correspond to modulationally unstable states with complex
characteristic velocities (\ref{305.4}).
}
\label{Fig3}
\end{figure}

It is useful to give another graphic representation of the rarefaction waves. From definition
(\ref{306.15}) of we Riemann invariants we find that they are constant along parabolas
\begin{equation}\label{315.2}
  \rho=-\frac1{2r}\left(\frac{u}2-r\right)^2
\end{equation}
in the $(u,\rho)$-plane, where $r$ is the value of the corresponding Riemann invariant.
If a rarefaction wave corresponds to $r_+=\mathrm{const}$, then both its left and right points must lie on
the same parabola shown in Fig.~\ref{Fig3}(a) by a blue line. These points can be represented as
crossing points of this blue parabola with other two parabolas that represent curves with constant
$r_-^L$ and $r_-^R$ and are shown by red lines. We have two pairs of ``left'' and ``right'' points
and obtain, consequently, two types of rarefaction waves described by the diagram Fig.~\ref{Fig1}(a).
These transitions $L_a\to R_a$ and $L_b\to R_b$ correspond to different signs in the formulas
(\ref{310.7}). As we see, both transitions give the growth of $\rho$ with increase of $\zeta$ in
agreement with the plots in Fig.~\ref{Fig2}(a).
In a similar way, the situations corresponding to the diagram Fig.~\ref{Fig1}(b) with
constant Riemann invariant $r_-$ are represented by the parabolas shown in Fig.~\ref{Fig3}(b).
Now transitions from the ``left'' points to the ``right'' ones give the growth of $\rho$ in
one case and its decrease in another case, as it is shown  in Fig.~\ref{Fig2}(b).
It is important to notice that according to Eq.~(\ref{310.7}) these transitions connect the
points with the same signs of $u$, that is they do not intersect the ordinate axis separating the
monotonicity regions. Thus, these
rarefaction waves connect the states belonging to the same regions of monotonicity of the
Riemann invariants.
In the next sections we shall generalize this graphical representation to other wave
structures what will be quite helpful in classification of possible wave structures evolving
from initial discontinuities.

Both solutions for $\rho_-$ can describe flow of liquid into vacuum---in case (\ref{310.7}) from left
to right and in case (\ref{311.1}) from right to left. It is worth noticing a curious particular solution
for $u^L=0$, when $r_+^0=0$, $r_-=2x/(3t)$ and we get $\rho=-x/(3t)$, $u=0$. It is easy to see that
dispersionless equations (\ref{305.7}) admit such a solution.

Considered here wave structures satisfy the conditions (a) $r_+^L = r_+^R$, $r_-^L<r_-^R$ or (b) $r_+^L<r_+^R$,
$r_-^L=r_-^R$. It is natural to ask,
what happens if we have the initial conditions satisfying opposite inequalities,
and to answer this question we have to consider the DSW structures.

\subsection{Cnoidal dispersive shock waves}

\begin{figure}[t] \centering
\includegraphics[width=10cm]{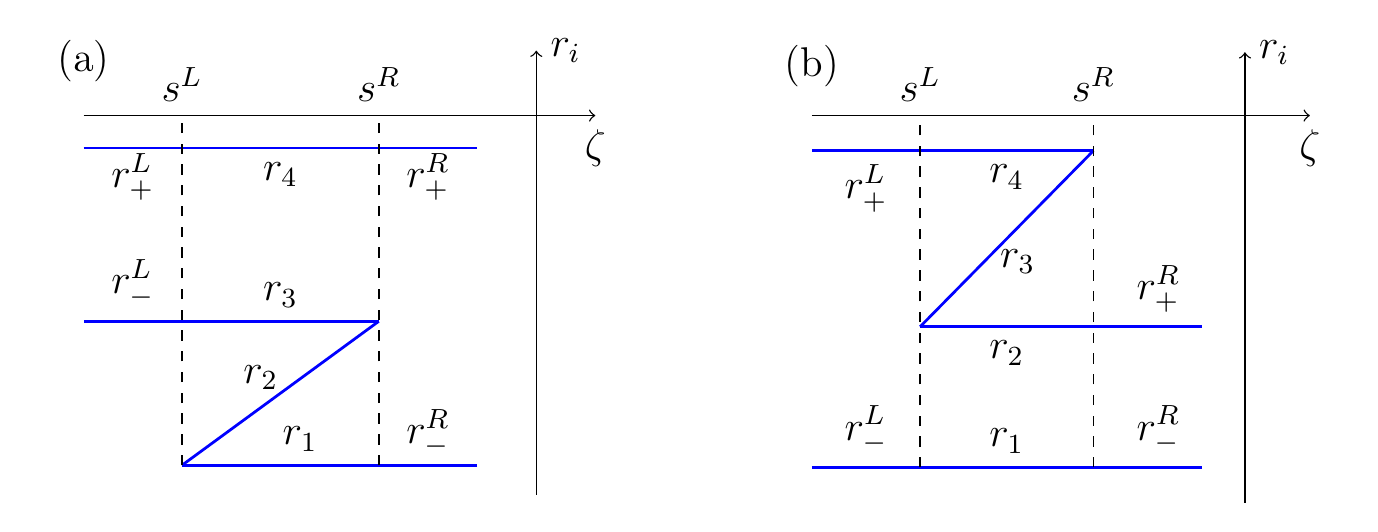}
\caption{Sketches of Riemann invariants for the DSW solutions of the DNLS equation: (a) $r_1,r_3,r_4$ are constant
and $r_2$ is determined by the equation $v_2=\zeta$; (b) (a) $r_1,r_2,r_4$ are constant
and $r_3$ is determined by the equation $v_3=\zeta$.
}
\label{Fig4}
\end{figure}

The other two possible solutions of Eqs.~(\ref{310.2}) are sketched
in Fig.~\ref{Fig4},
and they satisfy the boundary conditions (a) $r_+^L = r_+^R$,
$r_-^L > r_-^R$ or (b) $r_+^L > r_+^R$, $r_-^L = r_-^R$. In the
dispersionless approximation these multi-valued solutions are nonphysical.
However, following to Gurevich and Pitaevskii \cite{gp-73}, we can give them clear physical
sense by understanding $r_i$ as four
Riemann invariants of the Whitham system that describe evolution of a
modulated nonlinear periodic wave. Naturally, now $r_i$ are the self-similar solutions
of the Whitham equations (\ref{312.7}), that is of the equations
\begin{equation}\label{312.4}
  (v_i-\zeta)\frac{dr_i}{d\zeta}=0,\qquad i=1,2,3,4,
\end{equation}
which are obvious generalizations of (\ref{310.5}):
\begin{equation}\label{312.5}
  \eqalign{
  (a)\quad r_1=r_-^R,\quad r_3=r_+^R,\quad r_4=r_-^L,\quad v_2(r_-^R,r_2,r_-^L,r_+^L)=\zeta,\\
  (b)\quad r_1=r_-^R,\quad r_2=r_+^R,\quad r_4=r_+^L,\quad v_3(r_-^R,r_+^R,r_3,r_+^L)=\zeta,
  }
\end{equation}
where the last relations determine implicitly dependence of $r_2$ and $r_3$, correspondingly, on $\zeta$.
Sketches of these solutions are shown in Fig.~\ref{Fig4}. Velocities of the edges of the oscillatory zone whose
envelopes are described by the solutions of the Whitham equations are given by
\begin{equation}\label{312.6}
  \eqalign{
  (a)\quad &s^L=v_2(r_-^R,r_-^R,r_-^L,r_+^L)=2r_-^R+\frac{(r_+^L-r_-^L)^2}{2(r_+^L+r_-^L-2r_-^R)},\\
  &s^R=v_2(r_-^R,r_-^L,r_-^L,r_+^L)=\frac12(r_-^R+2r_-^L+r_+^L),\\
  (b)\quad &s^L=v_3(r_-^R,r_+^R,r_+^R,r_+^L)=\frac12(r_-^R+2r_+^R+r_+^L),\\
  &s^R=v_2(r_-^R,r_-^R,r_-^L,r_+^L)=2r_-^R+\frac{(r_+^L-r_-^L)^2}{2(r_+^L+r_-^L-2r_-^R)},
  }
\end{equation}
correspondingly.

\begin{figure}[t] \centering
\includegraphics[width=10cm]{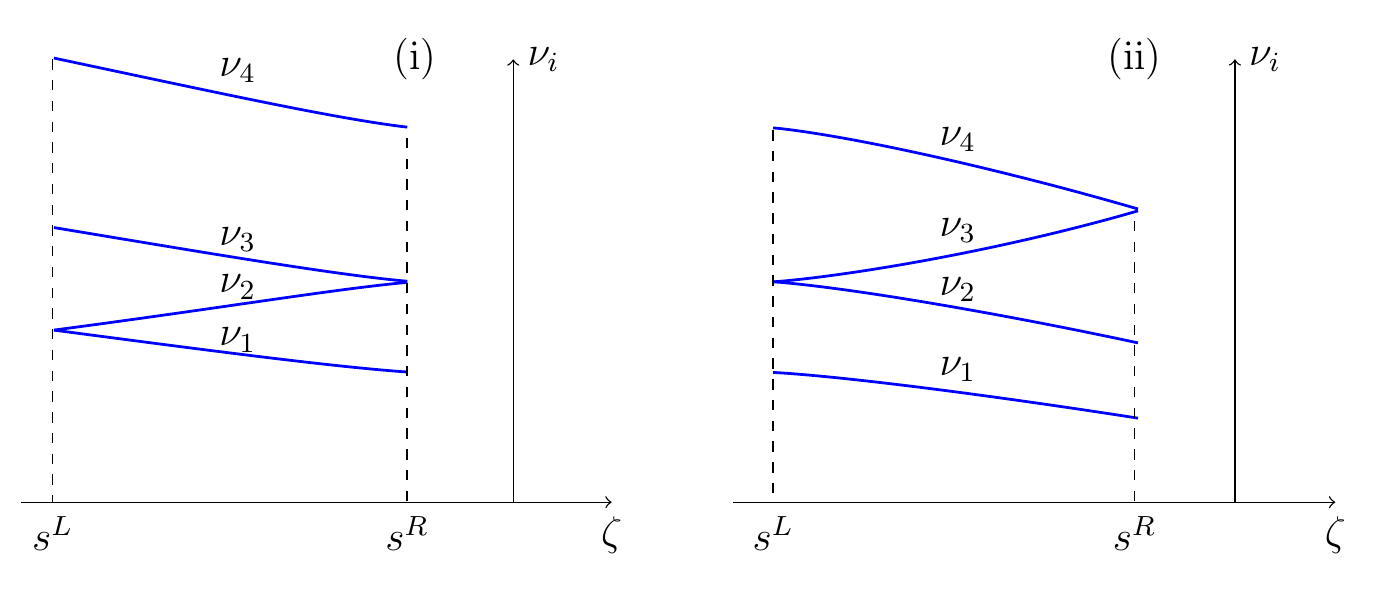}
\caption{Dependence of the parameters $\nu_i$ on $\zeta$ in self-similar solutions of the
Whitham equations.
}
\label{Fig5}
\end{figure}

If we substitute the solutions (\ref{312.5}) into formulae (\ref{312.3a}) and (\ref{312.3a}), then
we determine the dependence of the parameters $\nu_i$ on $\zeta$. There are two
possibilities shown in Fig.~\ref{Fig5}: the diagram Fig.~\ref{Fig4}(a) is mapped by both sets
of formulae (\ref{312.3a}) and (\ref{312.3b}) into the type Fig.~\ref{Fig5}(i), whereas the
diagram Fig.~\ref{Fig4}(b) is mapped by the formulae (\ref{312.3a}) into the type Fig.~\ref{Fig5}(ii)
and by the formulae (\ref{312.3b}) into the type Fig.~\ref{Fig5}(i).

\begin{figure}[t] \centering
\includegraphics[width=10cm]{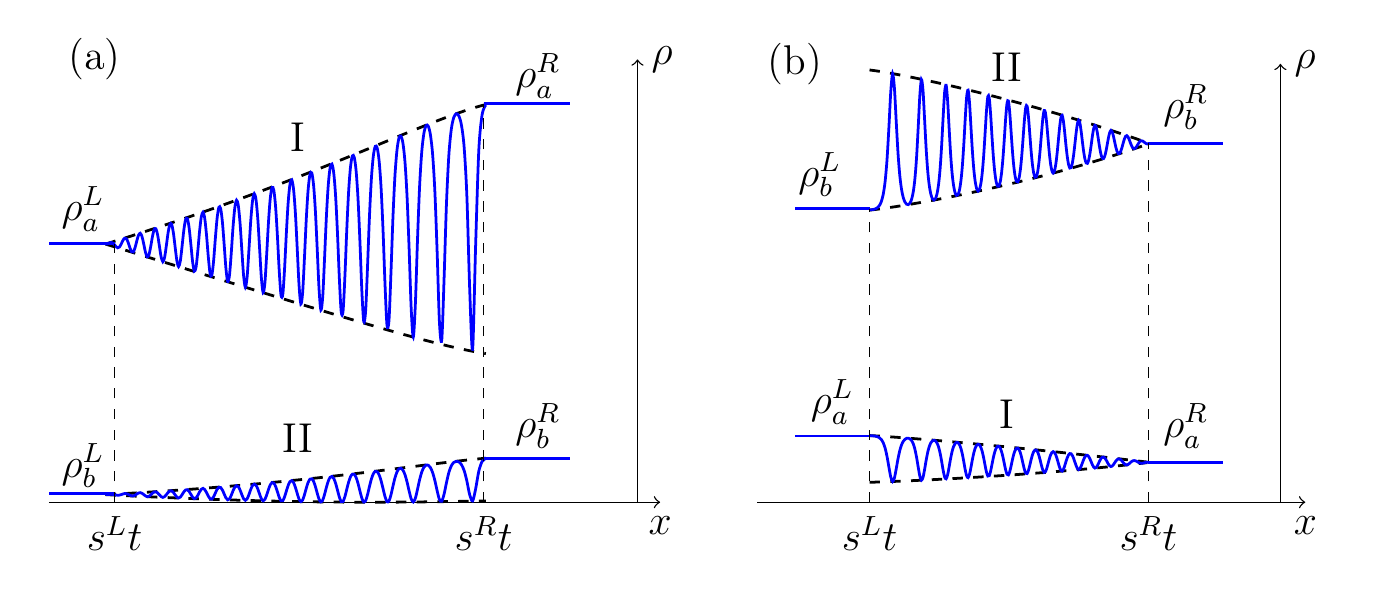}
\caption{Dispersive shock waves evolved from initial discontinuities for
(a) $r^L_+=r_4=r^R_+$, (b) $r^L_-=r_1=r^R_-$. The bold dashed lines indicate
envelopes of modulated nonlinear waves.
}
\label{Fig6}
\end{figure}

The solutions obtained here are interpreted as formation of
cnoidal dispersive shock waves evolving from initial discontinuities with such a type
of the boundary conditions. Indeed, Eq.~(\ref{eq20}) upon substitution of obtained $\nu_i$
yields the plots shown in Fig.~\ref{Fig6}(a)I,II and Fig.~\ref{Fig6}(b)II, whereas Eq.~(\ref{eq30})
yields the plot Fig.~\ref{Fig6}(b)I. We summarize the appearing possibilities in the following
list:
\begin{itemize}
  \item Fig.~\ref{Fig4}(a)\quad$\stackrel{(\ref{312.3a})}\longrightarrow$
  \quad Fig.~\ref{Fig5}(i)\quad$\longrightarrow$ \quad Fig.~\ref{Fig6}(a)\,\,\, plot I
  \item Fig.~\ref{Fig4}(a)\quad$\stackrel{(\ref{312.3b})}\longrightarrow$
  \quad Fig.~\ref{Fig5}(i)\quad$\longrightarrow$ \quad Fig.~\ref{Fig6}(a)\,\,\, plot II
  \item Fig.~\ref{Fig4}(b)\quad$\stackrel{(\ref{312.3a})}\longrightarrow$
  \quad Fig.~\ref{Fig5}(i)\quad$\longrightarrow$ \quad Fig.~\ref{Fig6}(b)\,\,\, plot II
  \item Fig.~\ref{Fig4}(b)\quad$\stackrel{(\ref{312.3b})}\longrightarrow$
  \quad Fig.~\ref{Fig5}(ii)\quad$\longrightarrow$ \quad Fig.~\ref{Fig6}(b)\,\,\, plot I
\end{itemize}
As we see, each solution of Whitham equations expressed in terms of Riemann invariants is mapped into
two different DSW structures which satisfy the boundary conditions compatible with given values of the
Riemann invariants. Such a behavior is typical for {\it non-convex} dispersive hydrodynamics
and has already been discussed in simpler situation of mKdV (Gardner) equation in Ref.~\cite{kamch-12}.

This two-valued connection of Riemann invariants with solutions in terms of physical variables is similar to the
situation described above for the rarefaction waves: the diagram Fig.~\ref{Fig1}(a) yields two
decreasing with $\zeta$ density distributions shown in Fig.~\ref{Fig2}(a) whereas the diagram
Fig.~\ref{Fig1}(b) yields decreasing and increasing distributions shown in Fig.~\ref{Fig2}(b).
These two types of wave structures will serve us as building blocks appearing in evolution of arbitrary
initial discontinuity. It is clear that these cnoidal DSWs are described by the same diagrams of
Fig.~\ref{Fig3} as the rarefaction waves, but with inverted ``left'' and ``right'' points. Hence, the
cnoidal DSWs still connect the states belonging to the same regions of monotonicity of the dispersionless
Riemann invariants. But there must be waves that connects the states at opposite sides of the $\rho$-axis
$u=0$ in the $(u,\rho)$-plane and they also appear as elementary building blocks
which are described by the self-similar solutions of the Whitham equations. We shall turn to
this type of waves in the next subsection.

\subsection{Trigonometric (contact) dispersive shock waves}

\begin{figure}[t] \centering
\includegraphics[width=10cm]{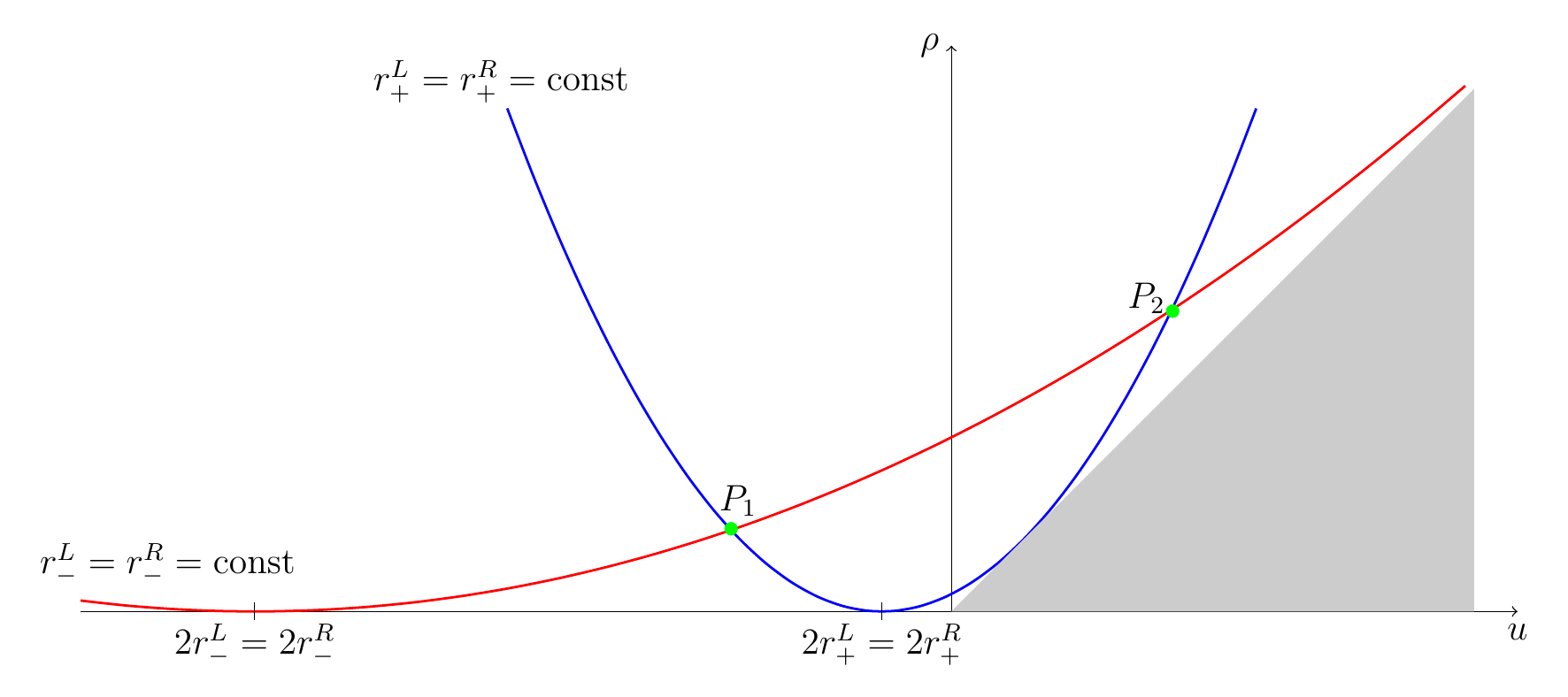}
\caption{Curves of constant Riemann invariants in the $(u,\rho)$-plane for transitions corresponding to
trigonometric dispersive shock waves. The boundary points have identical Riemann invariants $r_-^L=r_-^R$,
$r_+^L=r_+^R$.}
\label{Fig7}
\end{figure}

At first we shall consider the situation in which the Riemann
invariants have equal values at both edges of the shock, i.e., when
$r_-^L=r_-^R$, $r_+^L=r_+^R$. In this case we obtain a new type of wave
structure which we shall call a {\it contact dispersive shock wave} since
it has some similarity with contact discontinuities in the theory of
viscous shock waves (see, e.g., \cite{LL-6}).
For this situation, the parabolas corresponding to $r_-^L=\mathrm{const}$ and $r_-^R=\mathrm{const}$
in Fig.~\ref{Fig3}(a) coincide with each other and cnoidal DSWs disappear.
Instead, there appears the path connecting the identical left and right states
labeled by the crossing points of two parabolas as is shown in Fig.~\ref{Fig7}.
Such waves can arise only if the boundary points are located on the opposite
sides of the line $u=0$, i.e. in the different regions of monotonicity.

\begin{figure}[t] \centering
\includegraphics[width=8cm]{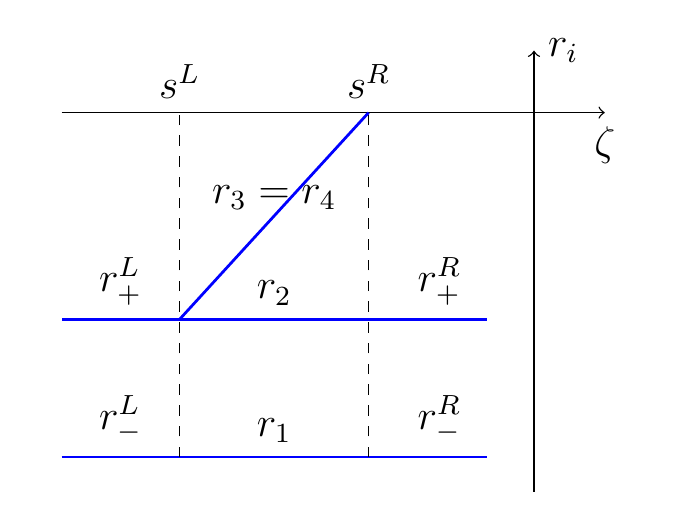}
\caption{Diagram of Riemann invariants for the trigonometric DSW solutions of the DNLS equation.}
\label{Fig8}
\end{figure}

Along the arc of the parabola connecting the points $P_1$ and $P_2$ the two biggest Riemann invariants
must be equal to each other, $r_3=r_4$, and at the left soliton edge they must equal to their boundary value
$r_3=r_4=r_2=r_+^L=r_+^R$. Hence, we arrive at the diagram of the Riemann invariants shown in Fig.~\ref{Fig8}.
Along this solution we have $m=0$ and the solutions of the Whitham equations is determined by the formula
\begin{equation}\label{317.1}
  v_3=v_4=2r_4+\frac{(r_+^L-r_-^L)^2}{2(r_+^L+r_-^L-2r_4)}=\zeta
\end{equation}
from which we obtain
\begin{equation}\label{317.5}
  r_3=r_4=\frac14\left[r_+^L+r_-^L+\zeta+\sqrt{(r_+^L+r_-^L-\zeta)^2+2(r_+^L-r_-^L)^2}\right].
\end{equation}
At the left soliton edge we have $r_4=r_+^L$ and at the right small-amplitude edge $r_4=0$.
Therefore Eqs.~(\ref{314.3}) yields velocities of the edges:
\begin{equation}\label{317.3}
  s^L=\frac32r_+^L+\frac12r_-^L,\qquad s^R=\frac{(r_+^L-r_-^L)^2}{2(r_+^L+r_-^L)}.
\end{equation}
The sign of the square root in Eq.~(\ref{317.5}) is chosen in such a way that this formula
gives $r_4=r_+^L$ at the left edge with $\zeta=(3r_+^L+r_-^L)/2$.

\begin{figure}[t] \centering
\includegraphics[width=12cm]{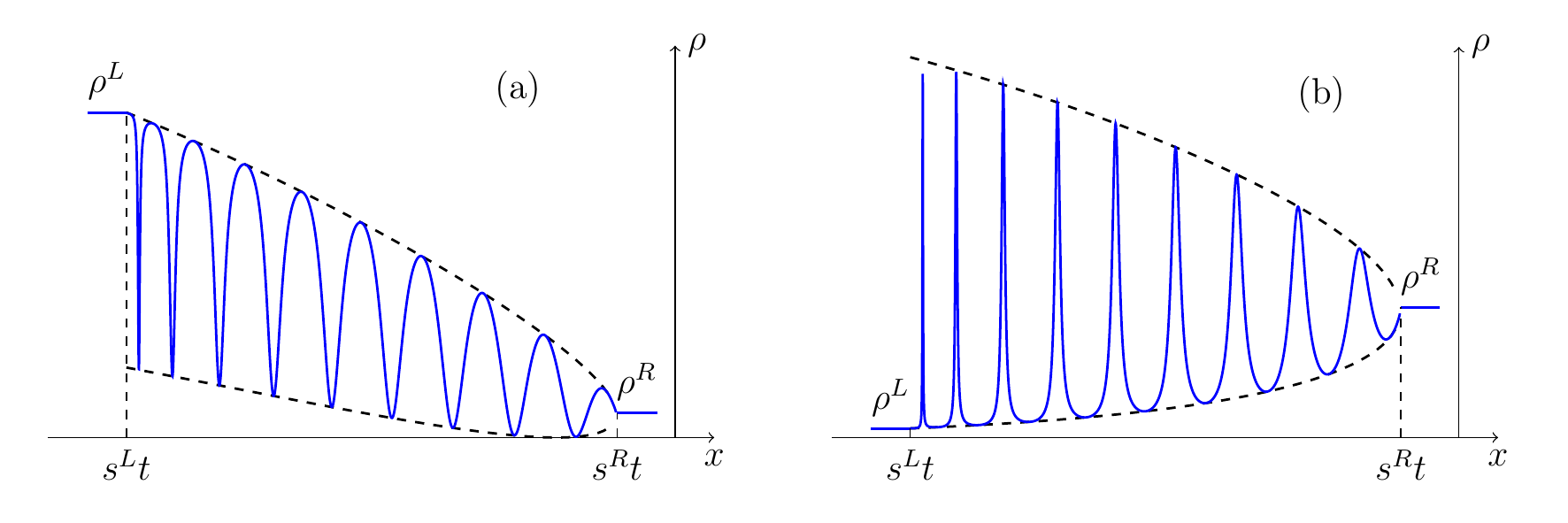}
\caption{Plots of $\rho$ the trigonometric DSW solutions of the DNLS equation,
(a) transition $P_2\to P_1$, (b) transition $P_1\to P_2$ in Fig.~\ref{Fig7}. Dashed thick lines
show the envelop functions obtained by solving the Whitham equations.}
\label{Fig9}
\end{figure}

As one can see from Eqs.~(\ref{312.3a}), in this case $\nu_3=\nu_4$ and $\nu$ oscillates in the
interval $\nu_1\leq\nu\leq\nu_2$. Then Eq.~(\ref{eq26}) yields the plot shown in Fig.~\ref{Fig9}(a)
with dark algebraic solitons at the left soliton edge. In case of Eqs.~(\ref{312.3b}) we have
$\nu_1=\nu_2$, hence $\nu$ oscillates in the interval $\nu_3\leq\nu\leq\nu_4$, and Eq.~(\ref{eq33})
yields the plot Fig.~\ref{Fig9}(b) with bright algebraic solitons at the soliton edge. Again the
same solution of the Whitham equations represented by a single diagram Fig.~\ref{Fig8} is mapped
into two different wave structures.

\subsection{Combined shocks}

\begin{figure}[t] \centering
\includegraphics[width=10cm]{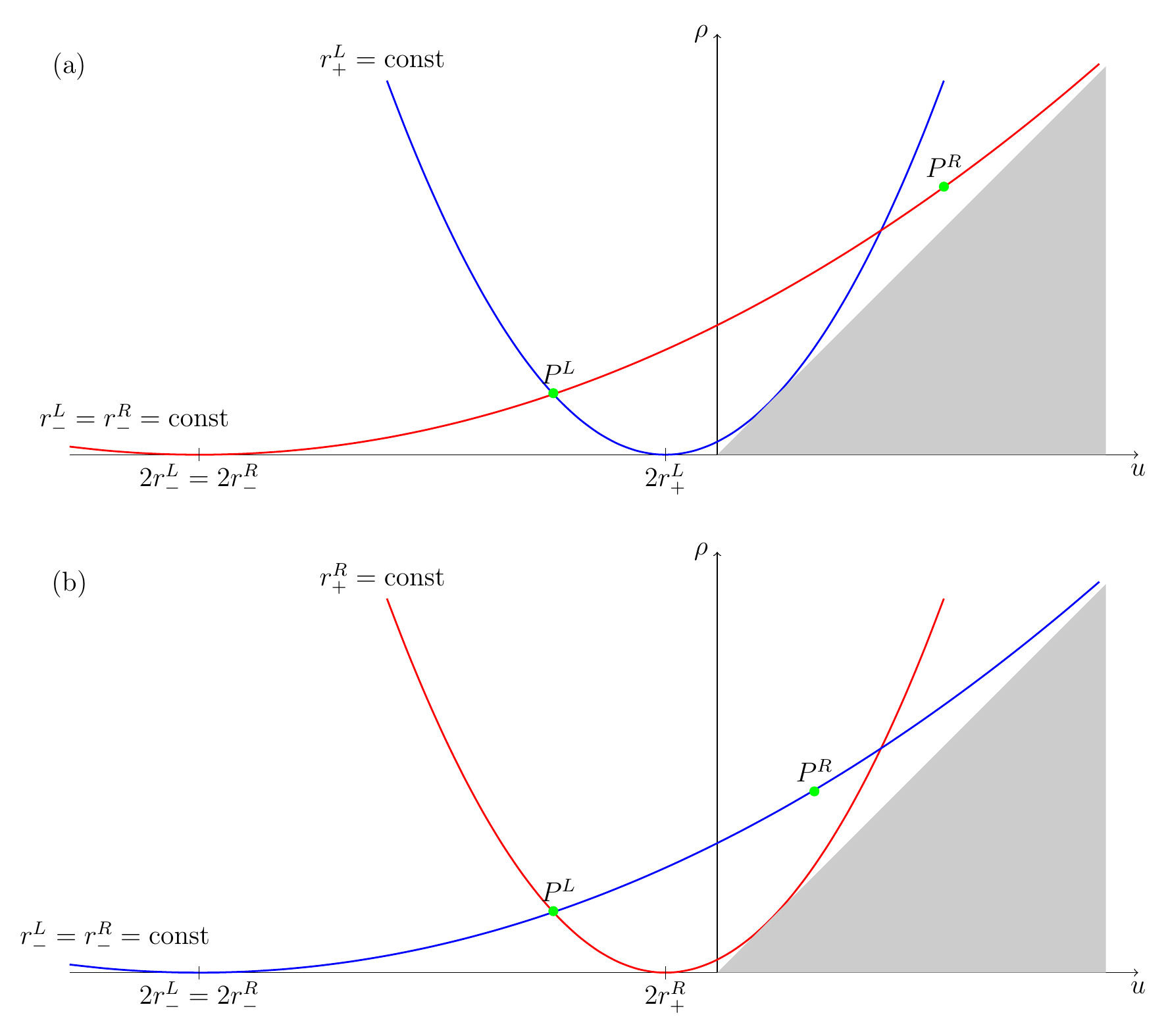}
\caption{Curves of constant Riemann invariants in the $(u,\rho)$-plane and transitions corresponding to
combined waves. Plot (a) represents the rarefaction waves with $r_+=\mathrm{const}$ combined with the cnoidal shock
and plot (b) corresponds to the trigonometric shock with $r_-=\mathrm{const}$ combined with the cnoidal shock.
}
\label{Fig10}
\end{figure}

\begin{figure}[t] \centering
\includegraphics[width=10cm]{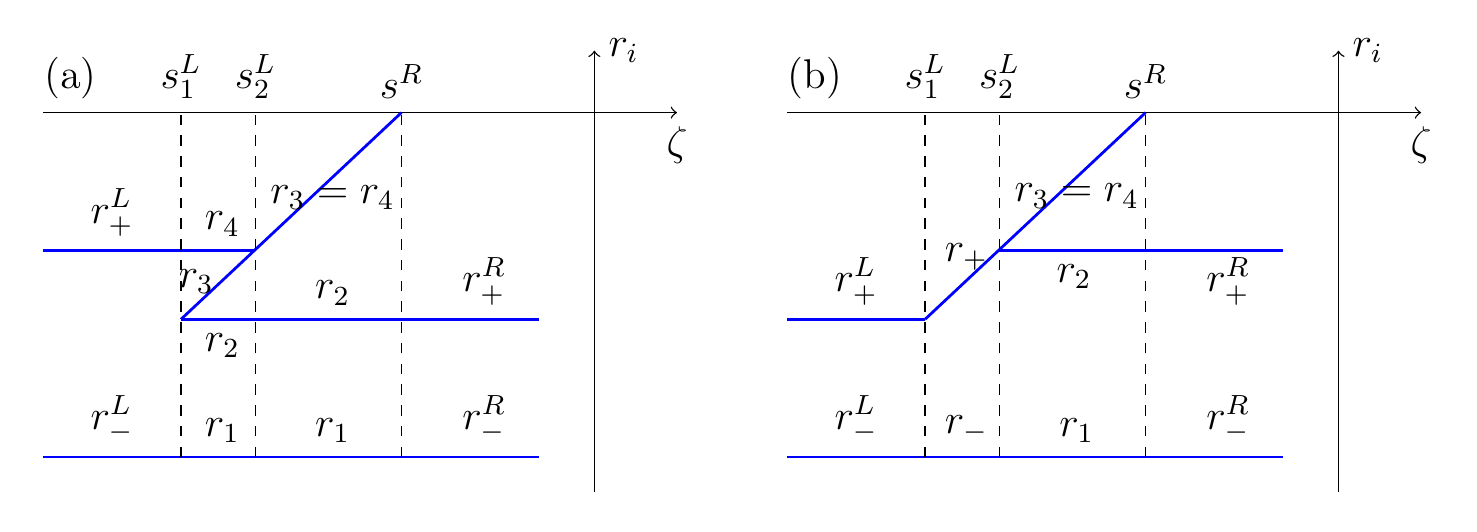}
\caption{(a) Diagram of the Riemann invariants for the cnoidal shock wave combined with the
trigonometric shock wave. (b)  Diagram of the Riemann invariants for the rarefaction wave combined with
the trigonometric shock wave.
}
\label{Fig11}
\end{figure}

\begin{figure}[t] \centering
\includegraphics[width=12cm]{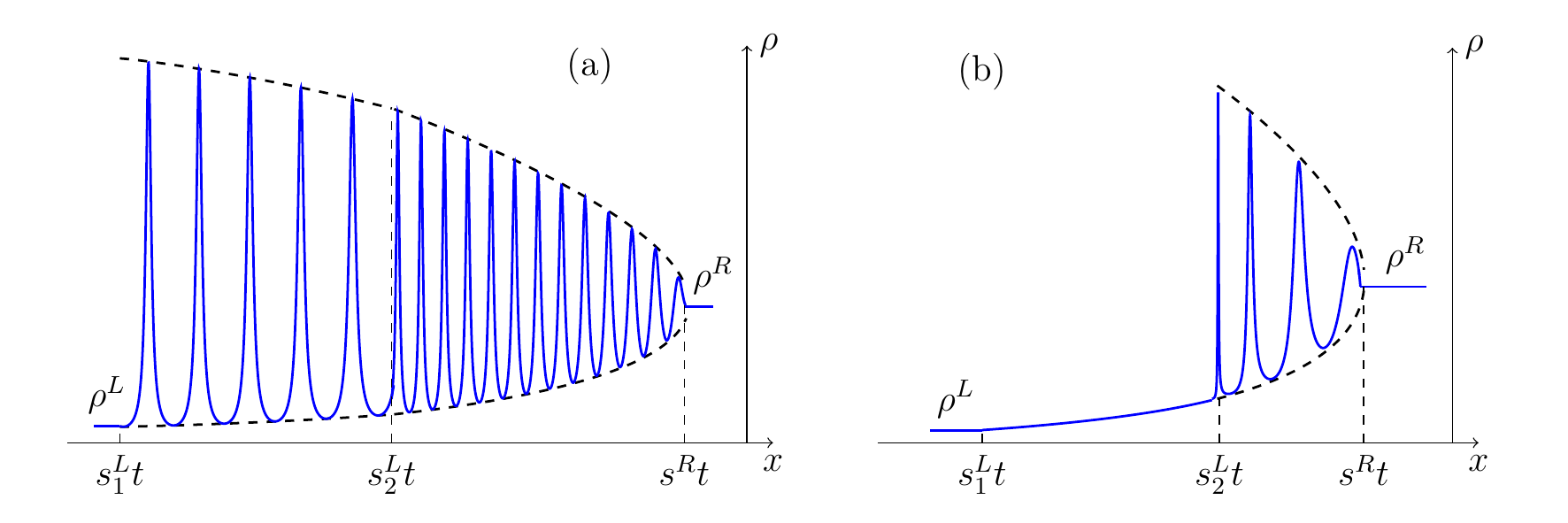}
\caption{(a) Combined shock wave consisting of the cnoidal and trigonometric dispersive shock waves. (b)
Combined shock wave consisting of the rarefaction wave and
the trigonometric shock wave.
}
\label{Fig12}
\end{figure}

Now we turn to the last elementary wave structures connecting two plateau states.
They can also be symbolized by single parabolic arcs between two points in the $(u,\rho)$-plane.
This type of paths is illustrated in Fig.~\ref{Fig10} and obviously it is a generalization
of the preceding structure. In this case, the boundary points are also located in
different monotonicity regions. One of the Riemann invariants still remains constant
($r_-^L=r_-^R$), however, the boundary values of the other Riemann invariant are different:
we have $r_+^L<r_+^R$ in case (a) and $r_+^L>r_+^R$ in case (b).
The corresponding diagrams of Riemann invariants are shown in Fig.~\ref{Fig11}.
As we see, in case (a) the oscillating region located between two plateaus consists of
two subregions---one with four different Riemann invariants, what corresponds to a
cnoidal DSW, and another one with $r_3=r_4$, what corresponds to a trigonometric DSW,
and there is no any plateau between them. Thus, this diagrams leads to a combined wave
structure of ``glued'' cnoidal and trigonometric DSWs. This structure is illustrated in
Fig.~\ref{Fig12}(a). At the soliton edge the cnoidal DSW matches with the left plateau and
the edge with $m=0$ it degenerates into the trigonometric shock. Velocities of the edge points
are equal to
\begin{equation}\label{318.1}
  \eqalign{s_1^L=v_3(r_-^L,r_+^R,r_+^R,r_+^L)=\frac12(r_-^L+2r_+^R+r_+^L),\\
  s_2^L=v_3(r_-^L,r_+^R,r_+^L,r_+^L)=2r_+^L+\frac{(r_+^R-r_-^R)^2}{2(r_+^R+r_-^R-2r_+^L)},\\
  s^R=v_3(r_-^L,r_+^R,0,0)=\frac{(r_+^R-r_-^R)^2}{2(r_+^R+r_-^R)}.
  }
\end{equation}

In a similar way, in case (b) we have a single trigonometric DSW
region glued with a rarefaction wave, as is shown in Fig.~\ref{Fig12}(b).
In this case the edge velocities are given by
\begin{equation}\label{318.2}
  \eqalign{
  s_1^L=\frac12(3r_+^L+r_-^L),\\
  s_2^L=\frac12(3r_+^R+r_-^R),\\
  s^R=v_3(r_-^L,r_+^R,0,0)=\frac{(r_+^R-r_-^R)^2}{2(r_+^R+r_-^R)}.
  }
\end{equation}
In both cases, the oscillatory wave is described by the formula (\ref{eq30}) or its
limit (\ref{eq33}) with oscillations of $\nu$ in the interval $\nu_3\leq\nu\leq\nu_4$.

Now, after description of all elementary wave structures arising in evolution of discontinuities
in the DNLS equation theory, we are in position to formulate the main principles of
classification of all possible wave structures.

\section{Classification of wave patterns}\label{sec8}

Classification of possible structures is very simple in the KdV equation case when any
discontinuity evolves into either rarefaction wave, or cnoidal DSW \cite{gp-73}.
It becomes more complicated in the NLS equation case \cite{eggk-95} and
similar situations as, e.g., for the Kaup-Boussinesq equation \cite{egp-01,cikp-17},
where the list consists of eight or ten structures, correspondingly, which can be
found after simple enough
inspection of available possibilities which are studied one by one. However, the situation
changes drastically when we turn to non-convex dispersive hydrodynamics: even in the case
of unidirectional Gardner (mKdV) equation we get eight different patterns (instead of two
in KdV case) due to appearance of new elements (kinks or trigonometric and combined
dispersive shocks), but these patterns can be labeled by two parameters only and
therefore these possibilities can be charted on a two-dimensional diagram.
In our present case the initial discontinuity (\ref{init}) is parameterized by
four parameters $u^L,\rho^L,u^R,\rho^R$, hence the number of possible wave patterns considerably
increases and it is impossible to present them in a two-dimensional chart.
Therefore it seems more effective to formulate the principles according to which
one can predict the wave pattern evolving from a discontinuity with given parameters.
Similar method was used \cite{ik-17,ikcp-17} in classification of wave patterns evolving
from initial discontinuities according to the generalized NLS equation and the Landau-Lifshitz
equation for easy-plane magnetics or polarization waves in two-component Bose-Einstein condensate.

\begin{figure}[t] \centering
\includegraphics[width=12cm]{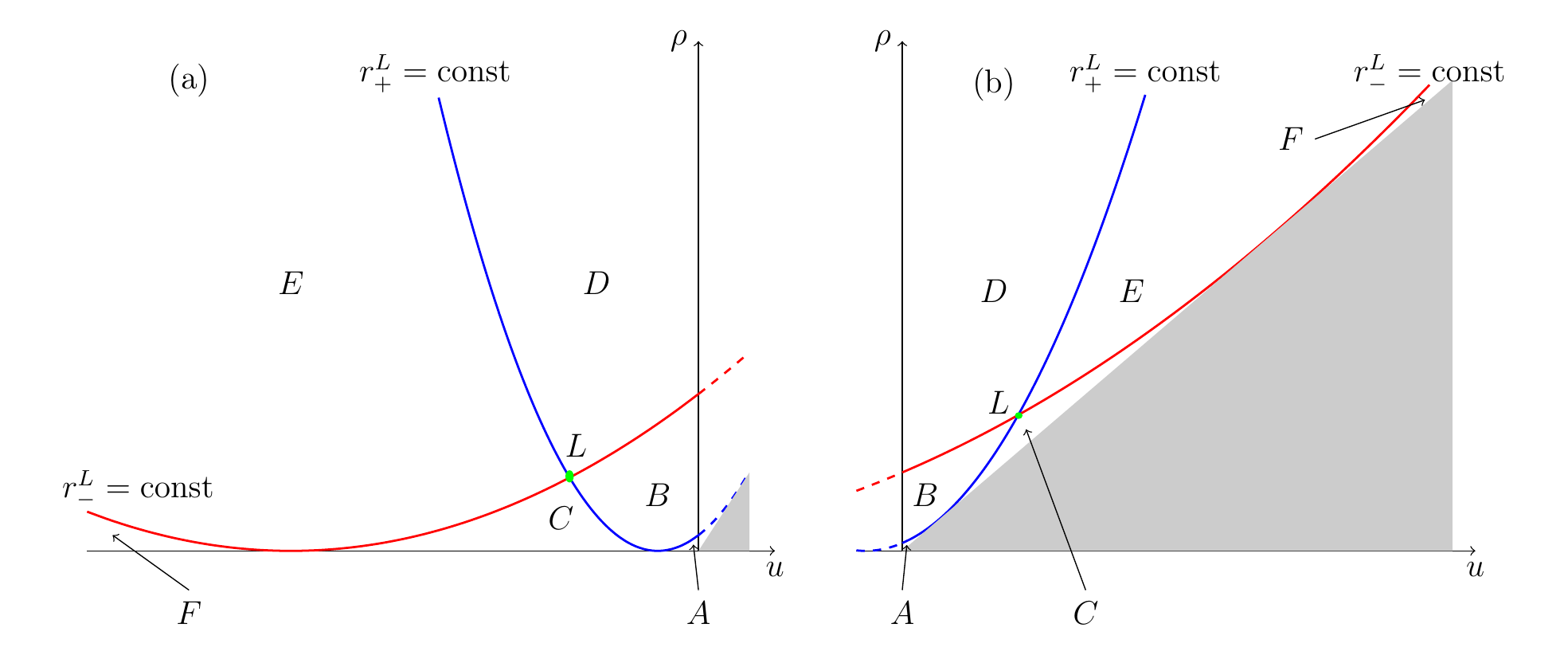}
\caption{(a) Domains corresponding to different wave structures for evolution on an initial
discontinuity whose both edges lie in the left monotonicity region $u<0$.
(b)
Domains corresponding to different wave structures for evolution on an initial
discontinuity whose both edges lie in the right monotonicity region $u>0$.
}
\label{Fig13}
\end{figure}

We begin with the consideration of the classification problem from the case
when both boundary points lie on one side of the axis $u=0$ separating two
monotonicity regions in the $(u,\rho)$-plane. At first we shall consider situation
when the boundary points lie in the left monotonicity region $u<0$.
We show in Fig.~\ref{Fig13}(a) the two parabolas corresponding to the constant
dispersionless Riemann invariants $r_{\pm}^L$ related with the left boundary state.
Evidently, they cross at some point $L(u^L,\rho^L)$ representing the left boundary. These two
parabolas cut the left monotonicity region into six domains labeled
by the symbols $A,B,\ldots,F$.  Depending on the domain, in which the point $R$ with
coordinates $(u^R,\rho^R)$, representing the right boundary condition, is located,
one gets one of the six following possible orderings of the left and
right Riemann invariants:
\begin{equation}\label{RiemannInequalities}
    \eqalign{
     \mbox{A}: \quad \la_-^R < \la_+^R < \la_-^L < \la_+^L, \\
       \mbox{B}: \quad \la_-^R < \la_-^L < \la_+^R < \la_+^L, \\
     \mbox{C}: \quad \la_-^L < \la_-^R < \la_+^R < \la_+^L, \\
        \mbox{D}: \quad \la_-^R < \la_-^L < \la_+^L < \la_+^R, \\
     \mbox{E}: \quad \la_-^L < \la_-^R < \la_+^L < \la_+^R, \\
        \mbox{F}: \quad \la_-^L < \la_+^L < \la_-^R < \la_+^R.
  }
\end{equation}
All these six domains and corresponding orderings yield six possible wave
structures evolving from initial discontinuities. Let us consider briefly each of them.

\begin{figure}
\centering
\includegraphics[width=4cm]{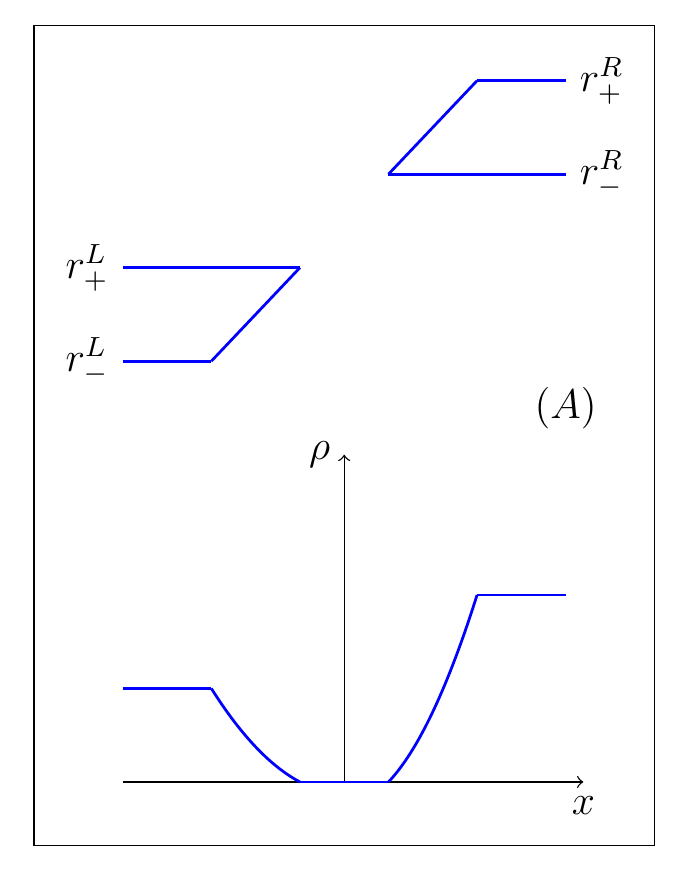}
\includegraphics[width=4cm]{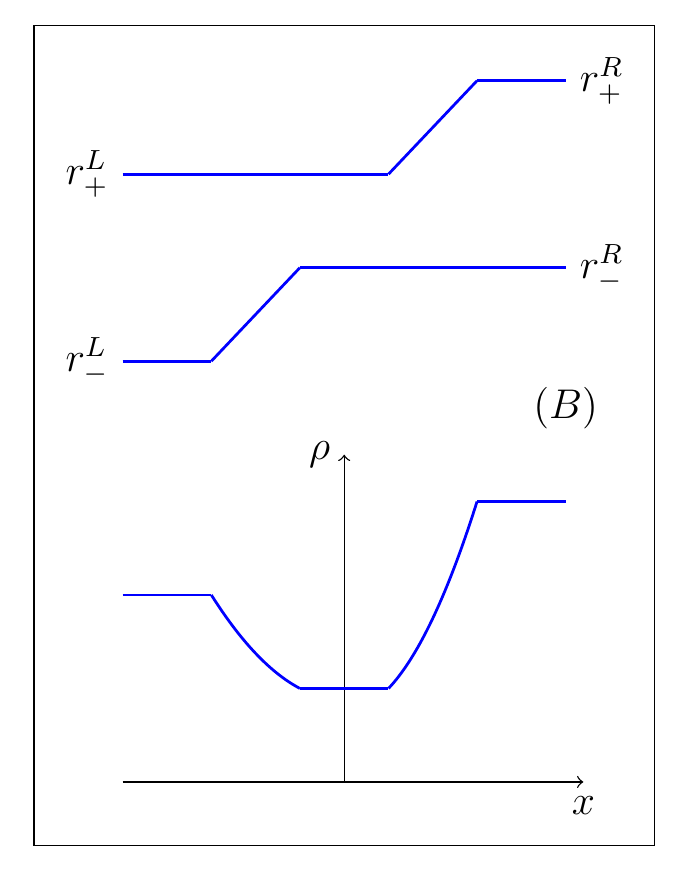}
\includegraphics[width=4cm]{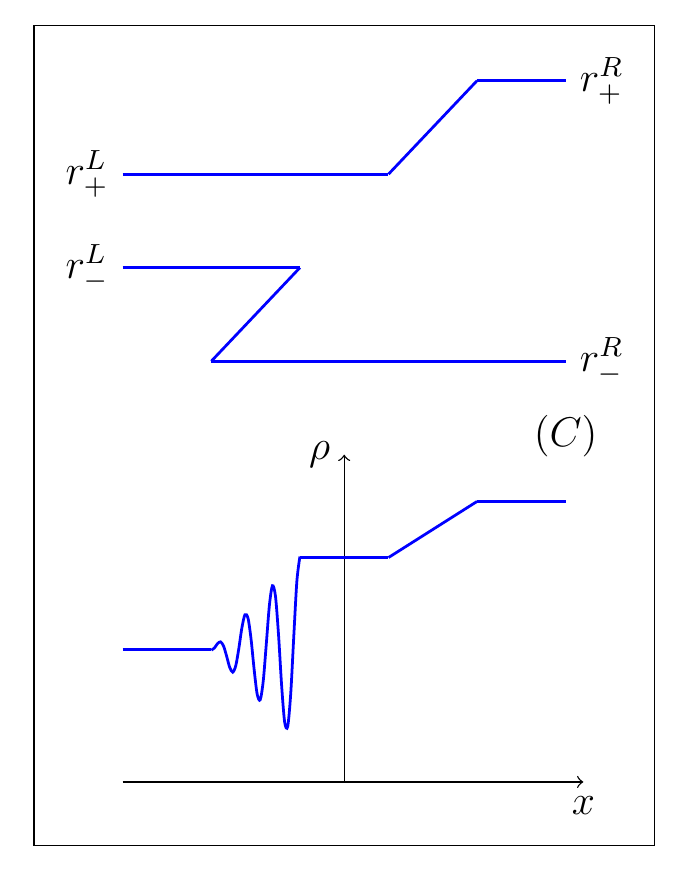}\\
\includegraphics[width=4cm]{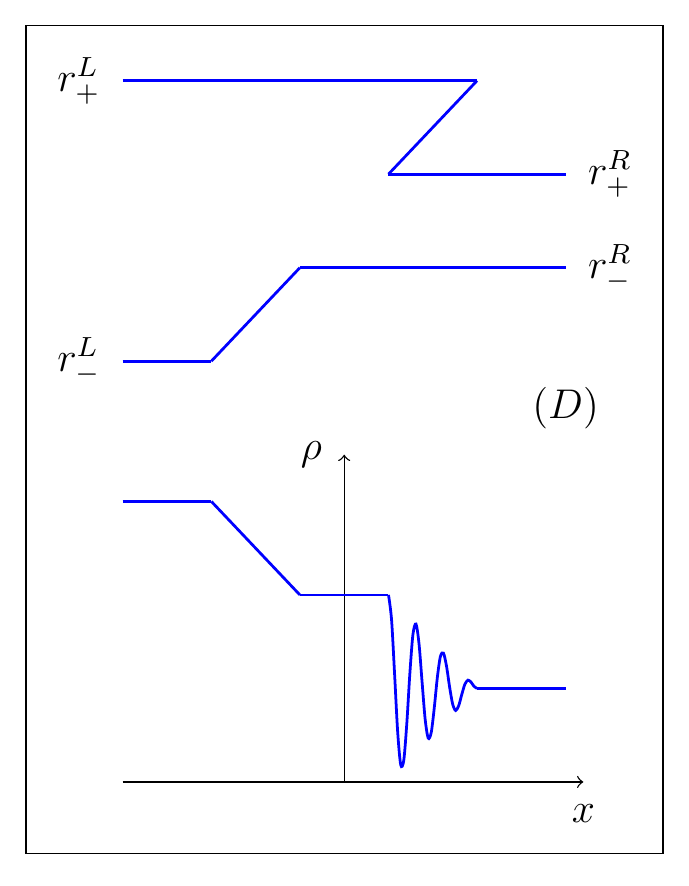}
\includegraphics[width=4cm]{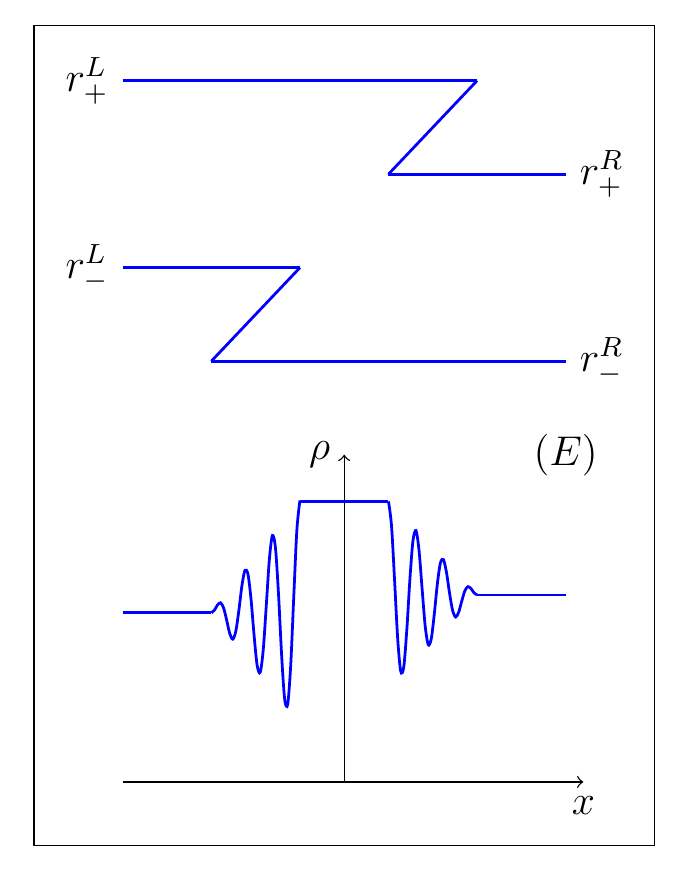}
\includegraphics[width=4cm]{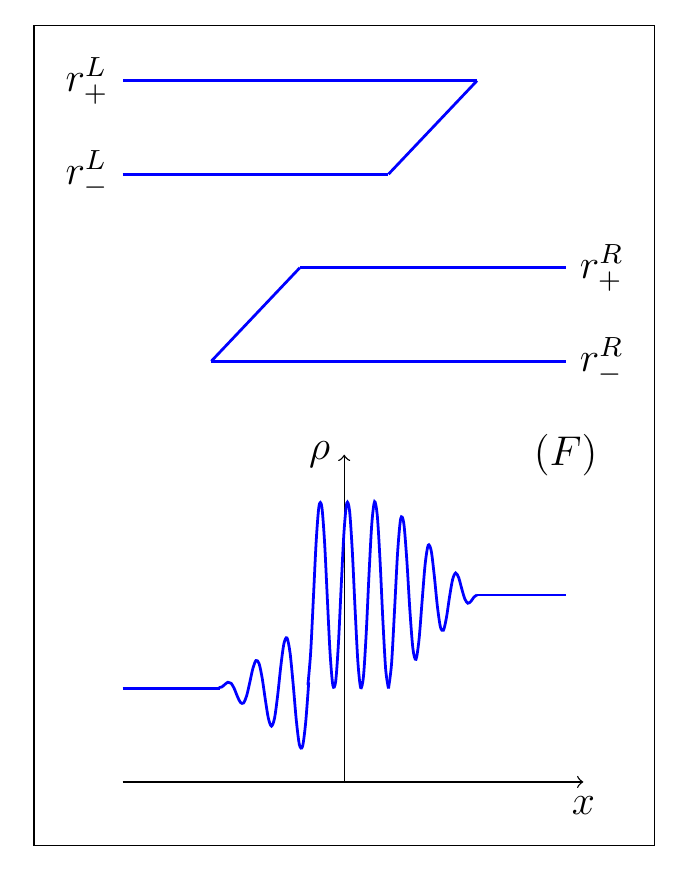}
\caption{Sketches of behavior of the Riemann invariants and of the corresponding wave
structures for six possible choices of the boundary conditions. }
\label{Fig14}
\end{figure}

In case (A) two rarefaction waves are separated by an empty region.
Evolution of Riemann invariants and sketch of wave structure
are shown in Fig.~\ref{Fig14}(A).

In case (B) two rarefaction waves are connected by a plateau whose
parameters are determined by the dispersionless Riemann invariants
$r_{\pm}^P$ equal to $r_-^P=r_-^R$ and $r_+^P=r_+^L$.
Here left and right ``fluids'' flow away from each other with small enough relative
velocity and rarefaction waves are able now to provide enough flux to create
a plateau in the region between them (see Fig.~\ref{Fig14}(B)).

In case (C) we obtain a dispersive shock wave on the left side of the structure,
a rarefaction wave on its right side and a plateau in
between (see Fig.~\ref{Fig14}(C)).

In case (D) we get the same situation as in the case (C),
but now the dispersive shock wave and rarefaction wave exchange their places
(see Fig.~\ref{Fig14}(D)).

In case (E) two DSWs are produced with a plateau between them.
Here we have a collision of left and right fluids (see Fig.~\ref{Fig14}(E)).

In case (F) the plateau observed in the case (E) disappears.
It is replaced by a nonlinear wave which
can be presented as a non-modulated cnoidal wave (see Fig.~\ref{Fig14}(F)).

The possible structures for this part of the $(u,\rho)$-plane
coincide qualitatively with the patterns found in similar classification problem
for the nonlinear Schr\"odinger equation \cite{eggk-95} and this case was already
studied in Ref.~\cite{gke-92}.

If we turn to consideration of the classification problem for the case
when both boundary points lie to the right of the line $u=0$, then
we get the diagram in the $(u,\rho)$-plane shown in Fig.~\ref{Fig13}(b).
We see that the parabolas cut again this right monotonicity region into
six domains. For this case the Riemann invariants can have
the same orderings (\ref{RiemannInequalities}) as in the previous case.
Depending on the location of the right boundary point in a certain domain,
the corresponding wave structure will be formed. Qualitatively
these structures coincide with those for the previous case.

At last, we have to study the situation when the boundary points lie on different
sides of the line $u=0$, that is in different monotonicity regions.
As we have seen in the previous section, in this case new complex structures, namely,
combined shocks, appear. It is easy to see that if the left boundary corresponds to
the point in the left monotonicity region, then we get again six wave patters, and if
it correspond to the point in the right monotonicity region, we get six more patterns,
twelve in total. In principle, they can be considered as generalizations of those
shown in Fig.~\ref{Fig14} with simple elements (rarefaction waves and cnoidal DSWs)
replaced by combined shocks. Instead of listing all possible patterns,
we shall formulate the general principles of their construction and illustrate them
by a typical example. This will provide
the method by which one can predict the wave pattern evolving from any given
initial discontinuity.

\begin{figure}[t] \centering
\includegraphics[width=12cm]{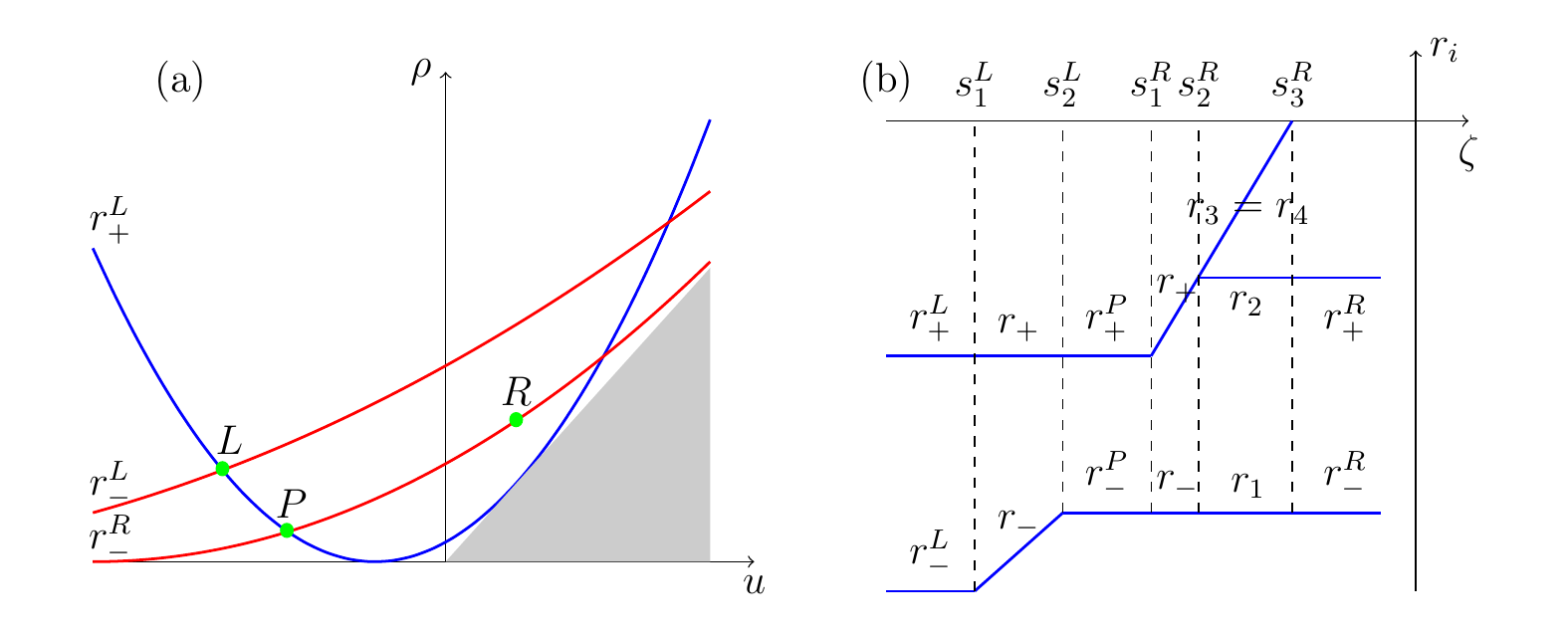}
\caption{(a) Arcs of parabolas of constant Riemann invariants that join the left $L$
and right $R$ states with plateau $P$ in between.
(b)
Diagram of Riemenn invariants corresponding to the path in the $(u,\rho)$-plane
shown in (a).
}
\label{Fig15}
\end{figure}

For given boundary parameters, we can construct the parabolas corresponding to
constant Riemann invariants $r_{\pm}^{L,R}$: each left or right pair of these
parabolas crosses at the point $L$ or $R$ representing the left or right
boundary state's plateau. Our task is to construct the path joining these two
points, then this path will represent the arising wave structure. We already know the answer
for the case when the left and right points lie on the same parabola, see, e.g.,
Fig.~\ref{Fig7}. If this is not the case and the right point $R$ lies, say,
below the parabola $r_-^L=\mathrm{const}$, see Fig.~\ref{Fig15}(a), then we can
reach $R$ by means of more complicated path consisting of two arcs of parabolas $LP$ and $PR$
joined at the point $P$. Evidently, this point $P$ represents the plateau between two waves
represented by the arcs. At the same time, each arc corresponds to a wave structure
discussed in the preceding section.
Having constructed a path from the left boundary point to the right one,
it is easy to draw the corresponding diagram of Riemann invariants.
To construct the wave structure, we use
the formulae connecting the zeroes $\nu_i$ of the resolvent with the Riemann
invariants $r_i$ and expressions for the solutions parameterized by $\nu_i$.
This solves the problem of construction of
the wave structure evolving from the initial discontinuity with given boundary conditions.
In fact, there are two paths with a single intersection point that join the left
and right boundary points and we choose the
physically relevant path by imposing the condition that velocities of edges of all
regions must increase from left to right.

\begin{figure}[t] \centering
\includegraphics[width=10cm]{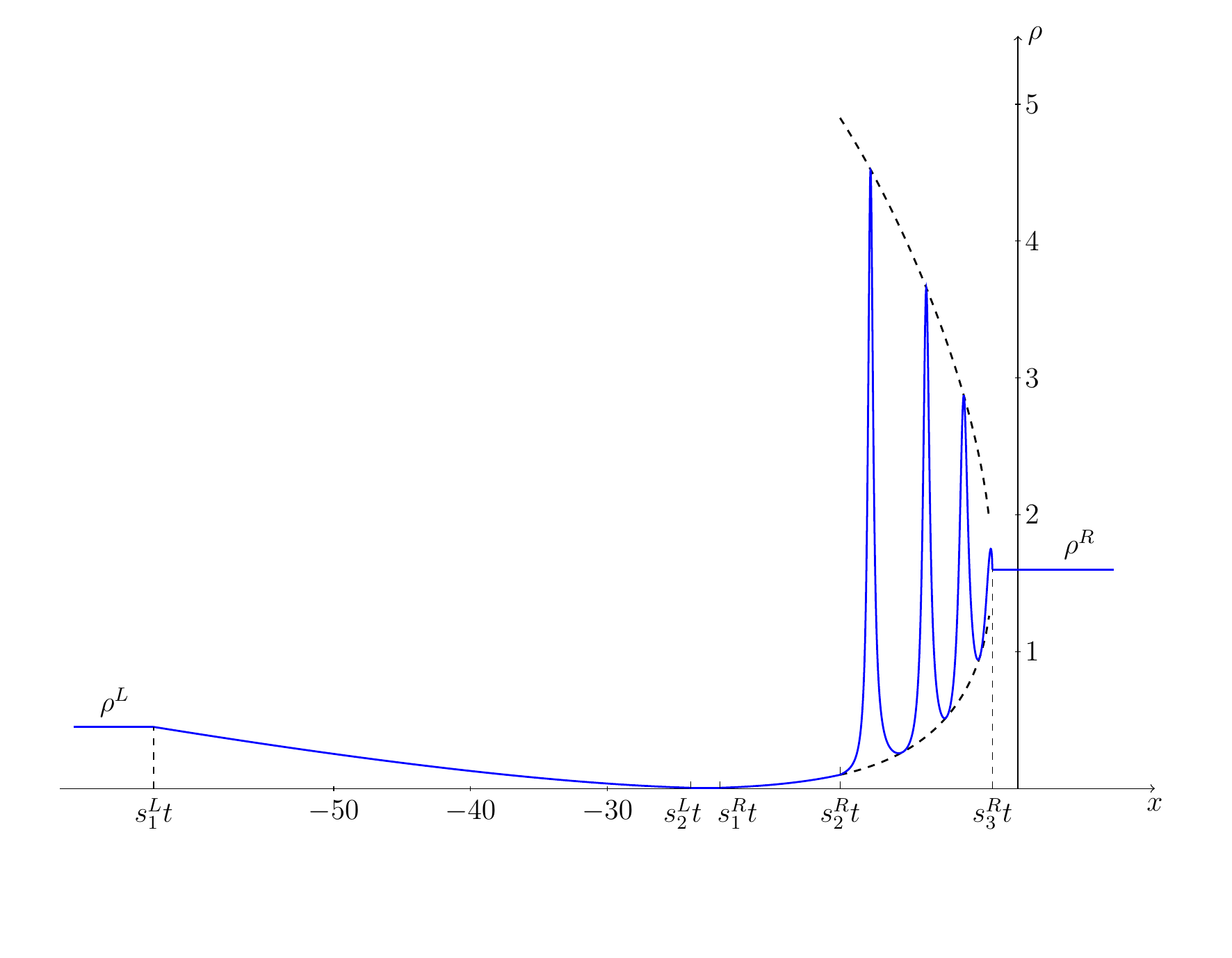}
\caption{Distribution of $\rho$ in evolutions of the initial discontinuity with
$\rho^L=0.45$, $u^L=-4$, $\rho^R=1.6$, $u^R=1.5$, what corresponds to the transition
$L\to R\in B(u>0)$ as in Fig.~\ref{Fig15}(a) and the diagram of the Riemann invariants
shown in Fig.~\ref{Fig15}(b).
The plateau between two rarefaction waves has the parameters $\rho^P=0.005$, $u^P=-2.27$.
The edge points have velocities $s_1^L=-6.32$, $s_2^L=-2.39$, $s_1^R=-2.18$,
$s_2^R=-1.3$, $s_3^R=-0.19$. The plot was calculated for $t=10$. Dashed thick lines show
envelope functions $\nu_3$ and $\nu_4$ given by Eqs.~(\ref{312.3b}) for the trigonometric
shock with $r_3=r_4$.
}
\label{Fig16}
\end{figure}

For example, let us consider the case $\rho^L=0.45$, $u^L=-4$, $\rho^R=1.6$, $u^R=1.5$
which corresponds to Fig.~\ref{Fig15}(a) with the transition $L\to R\in B(u>0)$.
In this case $r_-^L=-3.87$, $r_+^L=-1.035$, $r_-^R=-1.25$, $r_+^R=-0.45$ and we
see that the arc $PB$ of the parabola with $r_-^R=\mathrm{const}$ in the above transition
crosses the axis $u=0$ as is illustrated in Fig.~\ref{Fig15}(a).
Thus, we arrive at the diagram of Riemann invariants shown in Fig.~\ref{Fig15}(b).
Consequently, at the left edge we have a standard rarefaction wave (the arc $LP$ does not cross
the axis $u=0$) and at the right edge the combination of a trigonometric shock
with a rarefaction wave. Between these waves we get a plateau
characterized by the Riemann invariants $r_-^P=r_-^R$ and $r_+^P=r_+^L$. This plateau is represented by
a single point $P$ in Fig.~\ref{Fig15}(a). The rarefaction waves are described by the formulas
(\ref{310.7}) (left wave) and (\ref{311.1}) (right wave) with ``minus'' sight chosen in them.
The profile of the oscillatory wave structure can be obtained by substitution of the
solution of the Whitham equations
$$
r_1=r_-^R,\quad r_2=r_+^R,\quad v_3(r_-^R,r_+^R,r_3,r_3)=v_4(r_-^R,r_+^R,r_3,r_3)=\zeta,
$$
into Eq.~(\ref{eq33}) with $\nu_i$ given by Eqs.~(\ref{312.3b}). The velocities of the edge points are equal to
$$
\eqalign{
s_1^L=\frac12(r_+^L+3r_-^L),\quad s_2^L=\frac12(r_+^L+3r_-^R),\\ s_1^R=\frac12(3r_+^L+r_-^R),\quad
s_2^R=\frac12(3r-+^R+r_-^R),\quad s_3^R=\frac{(r_+^R-r_-^R)^2}{2(r_+^R+r_-^R)}.
}
$$
The resulting wave pattern is
shown in Fig.~\ref{Fig16}. It is easy to see, that it represents a deformation of the plot
Fig.~\ref{Fig14}(B): due to crossing the axis $u=0$ the right rarefaction waves
acquires a tail in the form of trigonometric DSW. It should be stressed that
appearance of such a tail is impossible in the theory of dispersive shock waves
in the NLS equation case.

In a similar way we can construct all twelve possible wave patterns for this type
on the boundary conditions.

\section{Conclusion}\label{sec9}

In this paper, we have developed the Whitham method of modulations for evolution of waves
governed by the DNLS equation. The Riemann problem of evolution of an initial
discontinuity is solved for this specific case of non-convex dispersive hydrodynamics.
It is found that the set of possible wave structures is much richer than in the convex case
(as, e.g., in the NLS equation theory) and includes, as structural elements, trigonometric
shock combined with rarefaction waves or cnoidal dispersive shocks.
Evolution of these trigonometric shocks is described by the
degenerate limits of the Whitham modulation equations. In the resulting scheme, one solution of the
Whitham equations corresponds to two different wave patterns, and this correspondence is provided by
a two-valued mapping of Riemann invariants to physical modulation parameters. Thus, the algebraic
resolvents introduced in Ref.~\cite{kamch-90a} for effectivization of periodic solutions of integrable
equations occurred to be crucially important also for establishing the relations between Riemann invariants
and modulation parameters of periodic solutions. 
To determine the pattern evolving from given discontinuity, we have
developed a graphical method which is quite flexible and was also applied to other systems with non-convex
hydrodynamics---generalized NLS equation for propagation of light pulses in optical fibers
\cite{ik-17} and Landau-Lifshitz equation for dynamics of magnetics with uniaxial easy-plane
anisotropy \cite{ikcp-17}.
The developed theory can find applications to physics of Alfv\'en waves in space plasma.

\ack
This work was partially supported by RFBR grant 16-01-00398. I am grateful to S K Ivanov
for useful discussions.

\end{document}